\pdfoutput=1
\documentclass[12pt]{article}

	\renewcommand\footnotemark{}

	\makeatletter
	\def\blfootnote{\xdef\@thefnmark{}\@footnotetext}
	\makeatother
\usepackage[printonlyused]{acronym}   
\usepackage{natbib}    

\usepackage{amssymb}  
\usepackage{amsmath}  
\usepackage{amsfonts}
\usepackage{appendix} 
\usepackage{graphicx} 
\usepackage{multicol} 
\usepackage{fullpage}  
\usepackage{rotating}  
\usepackage{epsfig}    
\usepackage{varioref}  
\usepackage{float}     
\usepackage{array}     
\usepackage{setspace}
\usepackage{threeparttable} 
\usepackage{longtable} 
\usepackage{supertabular} 
\usepackage{fancyhdr}  
\usepackage{color}
\usepackage{theorem}
\usepackage{layout}
\usepackage{makeidx}
\usepackage{acronym}
\usepackage{txfonts} 

	\usepackage{tikz} 
	\usetikzlibrary{calc}
	\usetikzlibrary{backgrounds} 
	\usetikzlibrary{plotmarks}

	\usepackage{xcolor}         
	\usepackage{multirow}       
	\usepackage{geometry}       
	\usepackage{threeparttable} 
	\usepackage{fullpage}       
	\usepackage{arydshln}       

\makeindex

\definecolor{myblue}{rgb}{0,.2,1}

\usepackage{hyperref}

\hypersetup{%
naturalnames=true,%
bookmarksnumbered=true,%
bookmarksopen=false,%
plainpages=true,%
colorlinks=true,%
urlcolor=myblue,
linkcolor=myblue,%
filecolor=myblue,%
citecolor=black,%
pdfpagemode=UseOutlines%
}%

\usepackage{palatino}

\DeclareMathOperator{\E}{E}

\DeclareMathOperator{\PrP}{Pr}

\bibpunct{(}{)}{;}{a}{}{;} 

\newcolumntype{M}{>{$}c<{$}}
\newcolumntype{R}{>{$}r<{$}}

\theoremstyle{plain}
\theorembodyfont{\normalfont}
\theoremheaderfont{\itshape}

\newtheorem{theorem}{Theorem}
\newtheorem{definition}{Definition}

\def\limfunc#1{\mathop{\rm #1}}%
	\newcommand{\mytitle}{An Economic Analysis of Privacy Protection and Statistical Accuracy as Social Choices}
	
	\newcommand{\myversion}{\today}  
	\newcommand{\mythanks}{Abowd and Schmutte acknowledge the support of Alfred P.\ Sloan Foundation Grant G-2015-13903 and NSF Grant SES-1131848. Abowd acknowledges direct support
	from NSF Grants BCS-0941226, TC-1012593.
	Any opinions and conclusions are those of the authors and do not represent the views of the Census Bureau, NSF, or the Sloan Foundation.
	We thank the Center for Labor Economics at UC--Berkeley and Isaac Newton Institute for Mathematical Sciences, Cambridge (EPSRC grant no.\ EP/K032208/1) for support and hospitality.
	We are extremely grateful for very valuable comments and guidance from the editor, Pinelopi Goldberg, and six anonymous referees. We acknowledge helpful comments from Robin Bachman, Nick Bloom, Larry Blume, David Card, Michael Castro, Jennifer Childs, Melissa Creech, Cynthia Dwork, Casey Eggleston, John Eltinge, Stephen Fienberg, Mark Kutzbach, Ron Jarmin, Christa Jones, Dan Kifer, Ashwin Machanavajjhala, Frank McSherry, Gerome Miklau, Kobbi Nissim, Paul Oyer, Mallesh Pai, Jerry Reiter, Eric Slud, Adam Smith, Bruce Spencer, Sara Sullivan, Salil Vadhan, Lars Vilhuber, Glen Weyl, and Nellie Zhao 
	along with seminar and conference participants at the U.S.\ Census Bureau, Cornell, CREST, George Mason, Georgetown, Microsoft Research--NYC, University of Washington Evans School, and SOLE.
	William Sexton provided excellent research assistance.
	No confidential data were used in this paper.
	Supplemental materials available at \url{http://doi.org/10.5281/zenodo.1345775}. The authors declare that they have no relevant or material financial interests that
relate to the research described in this paper.
	}
	\date{\myversion
	}

\acrodef{OPM}{Office of Personnel Management}
\acrodef{RAIS}{Rela\c{c}\~{a}o Anual de Informa\c{c}\~{o}es Sociais}
\acrodef{FOIA}{Freedom of Information Act}
\acrodef{UNECE}{United Nations Economic Commission for Europe}
\acrodef{FSS POS}{Federal Statistical System Public Opinion Survey}
\acrodef{SDL}{statistical disclosure limitation}
\acrodef{CNSS}{Cornell National Social Survey}
\acrodef{GSS}{General Social Survey}
\acrodef{CCD}{Common Core of Data}
\acrodef{BEA}{Bureau of Economic Analysis}
\acrodef{IRS}{Internal Revenue Service}
\acrodef{SOI}{Statistics of Income}
\acrodef{BLS}{Bureau of Labor Statistics}
\acrodef{CWHS}{Continuous Work History Sample}
\acrodef{SSA}{Social Security Administration}
\acrodef{SSN}{Social Security Number}
\acrodef{QCEW}{Quarterly Census of Employment and Wages}
\acrodef{CIPSEA}{Confidential Information Protection and Statistical Efficiency Act}
\acrodef{SPPE}{State Per-Pupil Expenditure}
\acrodef{RMSE}{root mean squared error}
\acrodef{PUMS}{public-use micro-data sample}
\acrodef{ACS}{American Community Survey}
\acrodef{RAPPOR}{Randomized Aggregatable Privacy-Preserving Ordinal Response}
\acrodef{DOE}{Department of Education}
	\renewcommand{\thesection}{\Roman{section}}
	\renewcommand{\thesubsection}{\Roman{section}.\Alph{subsection}}

	\title{\mytitle}
	\author{John M.\ Abowd and Ian M.\ Schmutte\thanks{\noindent 
	Abowd: U.S.\ Census Bureau HQ 8H120, 4600 Silver Hill Rd., Washington, DC 20233, and Cornell University, (email: \texttt{john.maron.abowd@census.gov)};
	Schmutte: Department of Economics, University of Georgia, B408 Amos Hall, Athens, GA 30602 (email: \texttt{schmutte@uga.edu}). \mythanks}}
	\date{\today \ \\ \ \\ Forthcoming in \emph{American Economic Review}}

\begin{document}

    \pagenumbering{gobble}
	\maketitle
	\newpage
	\pagenumbering{gobble}
    \begin{abstract}

%
%
\noindent
Statistical agencies face a dual mandate to publish accurate statistics while protecting respondent privacy. Increasing privacy protection requires decreased accuracy. Recognizing this as a resource allocation problem, we propose an economic solution: operate where the marginal cost of increasing privacy equals the marginal benefit. Our model of production, from computer science, assumes data are published using an efficient differentially private algorithm. Optimal choice weighs the demand for accurate statistics against the demand for privacy. Examples from U.S.\ statistical programs show how our framework can guide decision-making. Further progress requires a better understanding of willingness-to-pay for privacy and statistical accuracy.




    \end{abstract}
    \newpage
    \setcounter{page}{1}
    \pagenumbering{arabic}
    \onehalfspacing

	%


%
%

National statistical agencies collect information about the population and economy of a country directly from its people and businesses.
They face a dual mandate to publish useful summaries of these data while protecting the confidentiality of the underlying responses. These mandates are enshrined in law and official practices.\footnote{The \ac{CIPSEA} \citep{cipsea} and Census Act \citep{title13} obligate U.S.\ statistical agencies to protect confidentiality.}
For example, the Census Bureau is required by Article I of the U.S.\ Constitution to enumerate the population every ten years and by legislation to publish data from that census for the purpose of redrawing every legislative district in the country.\footnote{Under Public Law 94-171.
  } When providing these data, the Census Bureau is also subject to a legal prohibition against ``mak[ing] any publication whereby the data furnished by any particular establishment or individual ... can be identified.''\footnote{\citet[Section 9.a.2.]{title13}.}

The fundamental challenge posed in servicing this dual mandate is that as more statistics are published with more accuracy, more privacy is lost \citep{Dinur2003}.
Economists recognize this as a problem of resource allocation. We propose an economic framework for solving it.
Statistical agencies must allocate the information in their collected data between two competing uses: production of statistics that are sufficiently accurate balanced against the protection of privacy for those in the data. We combine the economic theory of public goods with cryptographic methods from computer science to show that social welfare maximization can, and should, guide how statistical agencies manage this trade-off.

\begin{figure}[th]
\centerline{\includegraphics[scale=0.65]{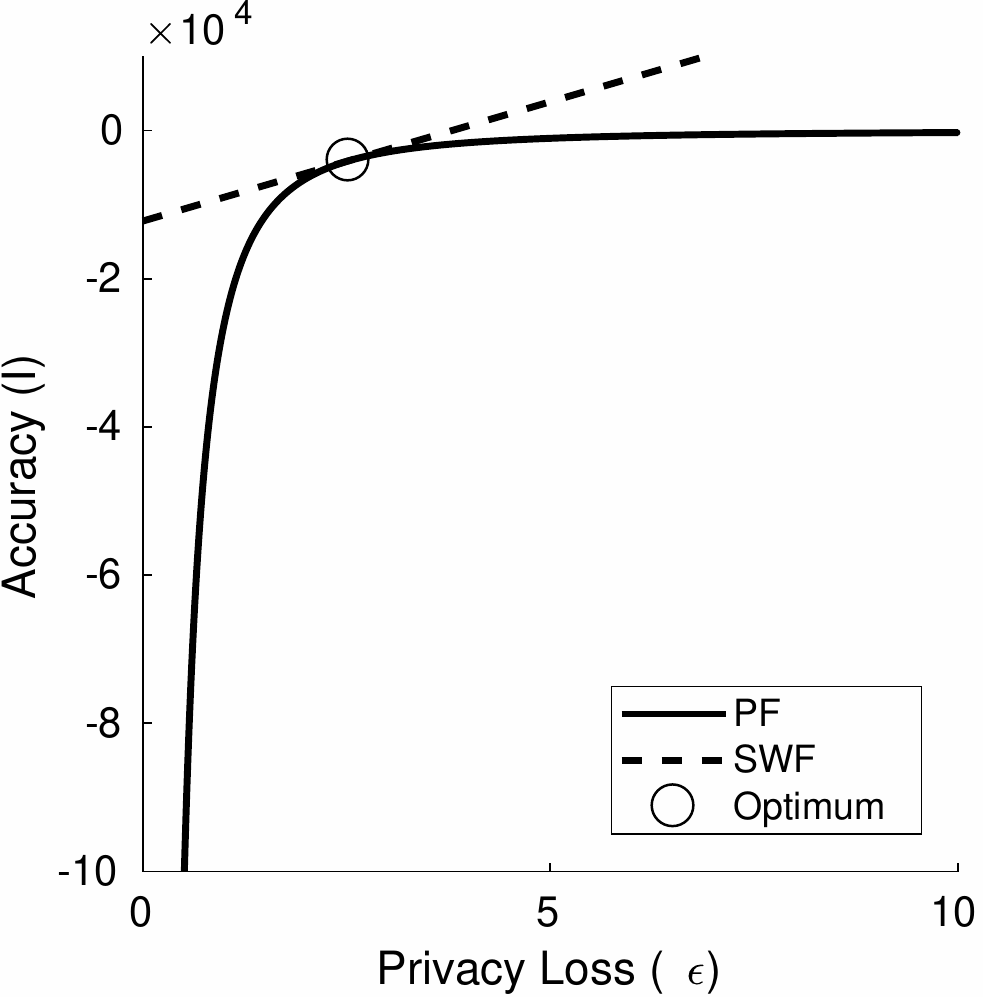}}
\caption{Solution to the Planner's Problem}
\label{fig:plannersprob}
\end{figure}
Figure \ref{fig:plannersprob} illustrates our adaptation of the approach proposed by \citet{samuelson:1954} for the efficient allocation of public goods. The horizontal axis measures \emph{privacy loss} parameterized by $\varepsilon$. The vertical axis measures \emph{accuracy}, parameterized by $I$. We define both concepts in detail in Section \ref{sec:preliminaries}. The line labeled \emph{PF} represents the production function, which describes feasible combinations of privacy loss and statistical accuracy available to the agency, given its endowment of data and known mechanisms for publishing. The line labeled \emph{SWF} is an indifference curve from the social welfare function defined in Section \ref{sec:preferences}. It describes aggregate preferences for privacy loss and accuracy. The optimal combination of privacy loss and accuracy is indicated by the open circle.
At that point, the marginal rate of transformation, which measures the cost of increased loss of privacy, matches the social willingness to accept privacy loss, both measured in units of increased statistical accuracy.

To date, this social choice framework has not been adapted to help guide statistical agencies in fulfilling their dual mandate.
Our key insight is that formal privacy systems developed in computer science can characterize the levels of privacy and accuracy available to a data custodian as a production function.
In our model, a statistical agency uses a known \emph{differentially private} mechanism to publish official statistics \citep*{Dworketal:2006,dwork2008differential,Dwork:Algorithmic:Book:2014}. Differential privacy measures privacy loss by the amount of information about any data record that is leaked when statistics are published. Differential privacy is very useful for describing the production technologies available to the statistical agency. However, formal privacy models shed no light on how to choose the right level of privacy or accuracy. This social choice question requires the tools of economics.\footnote{In related contributions from the electronic commerce literature, \citet{Hsu:EconomicEpsilon:IEEE:2014} and \citet{Ghosh:Auction:GEB:2015} model market-based provision of statistical summaries for private use by a single analyst in a setting where individuals can be directly compensated for their associated loss of privacy. In the context of these private transactions, \citet{Hsu:EconomicEpsilon:IEEE:2014} characterize an economic approach to setting the level of privacy loss. These models, like the formal privacy literature more generally, do not address the public-good nature of the published statistics and privacy protection offered by statistical agencies. The fundamental difference in our paper is the development of a social choice framework that accounts for all the benefits from the published statistics, not just those accruing to the explicit players as in the Hsu et al. and Ghosh and Roth models, and for all privacy-loss costs, not just the explicit losses for those who opt-in, because participation in our data collection is mandatory.}

Although the approach outlined in Figure \ref{fig:plannersprob} is familiar to economists, threats to privacy from publishing summary statistics may not be. Before discussing technical details, we discuss in Section \ref{sec:privacy} how differential privacy relates to practical, real-world concerns about confidentiality and data security. We give a plain-language interpretation of differential privacy and its relationship to the concerns individuals have about protecting whatever sensitive information has been stored in a database. We also discuss specific data breaches to highlight why statistical agencies and private data custodians are rapidly embracing differential privacy and other formal privacy-preserving data publication systems.

We model the publication of population statistics in Sections \ref{sec:preliminaries} and \ref{sec:production}. Section \ref{sec:preliminaries} focuses on the concepts we borrow from computer science.
In Section \ref{sec:production}, we interpret differentially-private publication mechanisms through the lens of producer theory. We give examples of mechanisms that yield closed, bounded, and convex production functions relating privacy loss and statistical accuracy. We define statistical accuracy as the expected squared error between the published value and the value that would be published in the absence of privacy protection. In Section \ref{sec:preferences}, we model preferences for privacy protection and accuracy as public goods and describe their optimal levels using a utilitarian social welfare function.

The issues raised in this paper are far from academic.
The threats to privacy inherent in the ``big data'' era have affected the policies governing statistical agencies.
In September 2017, the bi-partisan Commission on Evidence-based Policymaking recommended that agencies ``[a]dopt modern privacy-enhancing technologies ... to ensure that government's capabilities to keep data secure and protect confidentiality are constantly improving'' \citep[p.~2]{cep:promise:2017}.
That same month, the U.S.\ Census Bureau announced that it will use differential privacy, the leading privacy-enhancing technology, to protect the publications from its 2018 End-to-End Census Test and the 2020 Census.\footnote{
See
 \url{https://www.census.gov/about/cac/sac/meetings/2017-09-meeting.html} (cited on March 12, 2018).}
Our goal is to develop a principled framework for practical, empirically-driven, policy guidance regarding the balance of privacy protection and accuracy in modern statistical systems.

In Section \ref{sec:title_I}, we present an illustrative analysis of our framework applied to the allocation of federal funding to school districts under Title 1. In Section \ref{sec:applications}, we interpret several real-world problems in the U.S.\ statistical system using our social choice framework: the use of the PL94-171 tabulations to draw new legislative districts, the use of published inputs from the economic censuses to benchmark national accounts, the use of tax return data for tax policy modeling, and the publication of general-purpose micro-data. In each case, we describe the consequences of altering the weight on statistical accuracy versus privacy loss when facing an efficient frontier constraining the choices.

The allocation problem we study requires the perspective of economists---both in developing new theory and in connecting that theory to empirically-driven policy analysis.
Until now, our discipline has ceded one of the most important debates of the information age to computer science.\footnote{Privacy-preserving data analysis is barely known outside of computer science. A search for ``differential privacy'' in JSTOR's complete economics collection through December 2017 found five articles. The same query for statistics journals found six. A search of the ACM digital library, the repository for the vast majority of refereed conference proceedings in computer science, for the same quoted keyword found 47,100 results.}
In our conclusion, we draw attention to open questions that we hope will interest and inspire contributions from economists and other social scientists.


	\section{Key Concepts: Privacy and Accuracy}
    \label{sec:privacy}

Defining privacy in a rigorous but meaningful way is particularly challenging. To this end, we work with the concept of \emph{differential privacy}. This section explains how differential privacy relates to both data security and individual privacy. We also discuss our decision to measure statistical accuracy by expected squared error loss, and why this choice is not without loss of generality.

\subsection{Measuring Privacy}
\label{sec:measuring_privacy}
Differential privacy is a property of algorithms used to publish statistics from a confidential database. A differentially private algorithm guarantees that the published statistics will not change ``too much'' whether any observation from the confidential data is included or excluded.
The notion of ``too much'' is quantified by a parameter, $\varepsilon$, which measures the maximum difference in the log odds of observing any statistic across similar databases. For details, see Definition \ref{def:dif_priv}.

Differential privacy provides provable limits on an attacker's ability to re-identify individual records based on published statistics. Other methods of confidentiality protection are vulnerable to re-identification of large numbers of records through database reconstruction. We discuss database reconstruction, re-iden\-ti\-fi\-ca\-tion attacks, and the emergence of differential privacy in Section \ref{sec:breaches}.

Differential privacy can also guarantee individuals that their personal in\-for\-ma\-tion---secrets---will not be blatantly disclosed. Following computer science, we use the term \emph{semantic privacy} to refer to the latter sense of privacy protection.
Semantic privacy insures that what an attacker can learn about any person from published statistics does not depend ``too much'' on whether their data were used to compute the statistics.\footnote{
Statisticians will recognize the concept of semantic privacy as directly related to inferential disclosure \citep{Dalenius:Towards:1977,Goldwasser:Probabilistic:ACM:1982}.}
In Section \ref{sec:dp_semantics} we describe the conditions under which publication mechanisms guaranteeing $\varepsilon$-differential privacy also guarantee $\varepsilon$-semantic privacy.

Publication systems that are differentially private are \emph{closed under composition}, meaning that the cumulative privacy loss incurred by running multiple analyses on the same database can be computed from the privacy loss bounds on the component algorithms. Differentially private systems are \emph{robust to post-processing}, meaning that the privacy guarantee cannot be compromised by manipulation of the outputs, even in the presence of arbitrary outside information. Differentially private systems are  \emph{future proof}, meaning that their privacy guarantees do not degrade as technology improves or new data from other sources are published. Finally, differential privacy systems are \emph{public}, meaning that all the algorithms and parameters, except for the random numbers used in the implementation, can be published without compromising the privacy guarantee.
As argued in \citet{abowd:schmutte:BPEA:2015}, the public property of differential privacy is a major advantage over traditional \ac{SDL} because it allows for verification of the privacy protection and correct inferences from the published statistics.

There are caveats. While the differential privacy bound is useful for characterizing production possibilities, its interpretation is different from other economic variables that are conceptually well-defined but not precisely measurable. Differential privacy is a nominal bound on the worst-case privacy loss faced by any individual. Realized privacy loss depends on the actual data, the published statistics, and external information that is, or may become, available.
We characterize the statistical agency as choosing how to operate a system that guarantees a bound on everyone's privacy loss of $\varepsilon$, and allows verifiable accounting of compliance with that guarantee.

The worst-case bound in the definition of differential privacy is necessary for closure under composition. Closure under composition allows the custodian to compute global privacy loss as $\varepsilon$ for the entire set of published statistics. Closure under composition also preserves the non-rival public-good property of differential privacy protection. The global privacy protection parameterized by $\varepsilon$ is therefore the relevant public good that an economic analysis must allocate, not the loss of privacy experienced by a particular individual after publication of a particular statistic.

\subsection{Data Security and Privacy}
It may not be immediately obvious to applied researchers how publishing summary data can threaten privacy.
To motivate our analysis, we describe real-world violations of data security and situate them within a framework that reflects the common understanding of confidentiality protection in economics, statistics and computer science.
We defend the premise that blatant breaches of data security---``hacks''---are unacceptable.
The key insight from computer science is that publication of summary statistics leaks the same kind of private information as a breach.
Differential privacy has emerged as a focal paradigm because it can provably circumscribe such leakages.

\subsubsection{Motivating Examples}
\label{sec:motivating}
For an example of the harms of breaching data security, one need look no further than the Census Bureau's activities in WWII, releasing small-area data for the purposes of Japanese internment to the Army and providing individual records of the Japanese Americans in Washington, DC to the Secret Service for purposes of surveillance \citep{jones:2017}. Statutory barriers now prevent explicit data-sharing of this sort, and the U.S.\ Census Bureau staunchly guards those barriers \citep{Census:2002}. The Secretary of Commerce's March 26, 2018 direction to include a question on citizenship on the 2020 Census in support of providing block-level data on the citizen voting-age population by race and ethnicity makes the question of how best to protect the confidentiality of the micro-data in these publications even more salient.\footnote{\citet*{commerce.gov_2018}.}  There remains a threat that detailed data on a sensitive population could be accidentally shared by publishing so much summary information that the underlying data can be reverse-engineered. Therefore, statistical publications should guard against \emph{database reconstruction}.

Privacy is also threatened by \emph{data re-identification}. In 2006, Netflix ran a contest to improve its recommendation system. Netflix released data from its database of the ratings histories of its subscribers. To protect their users' privacy, Netflix removed direct identifiers and then released only a random sample of ratings. These efforts were not sufficient. \citet{Narayanan:2008:RDL:1397759.1398064} successfully re-identified a large share of the users in the Netflix Prize data by probabilistically matching to data scraped from \texttt{IMDb.com}, a crowd-sourced movie review site. This attack harmed the re-identified users because the attacker learned about movies they had rated in Netflix that they had not publicly rated in IMDb.\footnote{\citet{Garfinkel:Deidentification:NIST:2015} provides a more exhaustive overview of re-identification attacks.}


\subsubsection{From Data Breaches to Differential Privacy}
\label{sec:breaches}
To understand why differential privacy has captured the attention of statisticians, computer scientists, and economists, we elaborate on the concepts of \emph{database reconstruction} and \emph{data re-identification}.\footnote{We prefer the term \emph{statistical disclosure limitation} to \emph{anonymization} or \emph{de-identification}. All three are synonymous. We prefer the term \emph{re-identification} to \emph{de-anonymization}, but they also are synonymous.}
These concepts summarize key ideas from the literatures on formal privacy and \ac{SDL}, a complete review of which is beyond the scope of this paper.\footnote{See \citet{duncanetal2011statistical} for a comprehensive review of the \ac{SDL} literature. \citet{Heffetz2014} summarize differential privacy for economists.}


A \emph{reconstruction attack} is an attempt to build a record-by-record copy of a confidential database using only statistics published from it. This record-level image reproduces all the variables that were used in any published statistic to some level of accuracy. The accuracy depends on how many linearly independent statistics were published.
Call the list of variables subject to reconstruction ``list $A$''. Most published statistics do not include exact identifiers like name, address, \ac{SSN}, employer identification number, or medical case identifier. Call this list of identifiers ``list $B$''.

A \emph{re-identification attack} involves the linkage of records containing variables on list $B$ to records with variables on list $A$, either deterministically or probabilistically.
Records that link are called \emph{putative re-identifications}. They are unconfirmed claims that some entity on list $A$ belongs to a specific record associated with variables on list $B$.
A reconstruction attack abets a re-identification attack by facilitating the linkage of external information from list $B$. It allows the attacker to compare the data on the reconstructed variables from list $A$ with similar variables in external databases that also contain some variables from list $B$, generating putative re-identifications.

Our use of differential privacy is motivated by the \emph{database reconstruction theorem} due to \citet{Dinur2003}, which proves that conventional \ac{SDL} is inherently non-private. If the data custodian publishes many linearly independent statistics, then the confidential database can be reconstructed up to a very small error.
Reconstructing the confidential variables is
 a data breach.
Methods vulnerable to database reconstruction attacks are called \emph{blatantly non-private}.\footnote{Legally, whether a reconstruction-abetted re-identification attack constitutes an actionable data breach depends upon details of the attack. The U.S.\ government defines a data breach with reference to
personally identifiable information \citep[p. 8]{OMB:2014} as ``information that can be used to distinguish or trace an individual's identity, either alone or when combined with other information that is linked or linkable to a specific individual.'' The actionable consequences of a breach by this definition are implemented in agency policies \citep[e.g.][]{ds022:2014}.}

The database reconstruction theorem sounded the death knell for the \ac{SDL} methods typically used by statistical agencies.
Statistical agencies consider correct re-identification to be inherently problematic \citep[p. 103]{spwp22}.
Re-iden\-ti\-fi\-ca\-tion attacks can be facilitated by using the information from successful reconstruction attacks on public-use tables and micro-data.
Meanwhile, the amount of auxiliary information available to validate re-identification is rapidly increasing \citep{Garfinkel:Deidentification:NIST:2015}. So is the computing power and algorithmic sophistication needed to carry out these attacks.



\subsubsection{Reconstruction and Re-identification in Economic Data}
\label{sec:reconstruction}
We consider several examples of database reconstruction and re-identification risks in economic data. One such breach occurred when the \ac{CWHS} was released to researchers.
Its designers were so explicit about how the digits in the \ac{SSN} were used to construct the sample that a researcher could partially reconstruct valid SSNs known to be in the released data \citep{perlman:mandel:1944, 10.2307/41832028, perlman:1951, smith:1989}. The researcher could then re-identify individuals using the other variables.
The data were deemed to be in violation of the Tax Reform Act of 1976, and the \ac{CWHS} files were recalled \citep[p. 56]{buckler:1988}.\footnote{These are the data used by \citet{doi:10.2307/2118478}.
}

Researchers at the Census Bureau have acknowledged that their tabular publications may be vulnerable to database reconstruction attacks \citep{abowd:fcsm:2016, csac:2017, Abowd:JPC:2017}. This recognition is based, in part, on careful analysis of the statistics in all of the publications from the 2010 Census. More than 7.7 billion unique statistical summaries were published from 309 million persons---25 statistics per person. Each linearly independent statistic is one equation in a system that can be used for a record-level database reconstruction. For more details, see Appendix \ref{app:reconstruction_details}.

The county tables in the \ac{QCEW} released by the \ac{BLS} are also vulnerable to a reconstruction attack.\footnote{\citet*{Holan2010} published this attack. Toth is a BLS statistician.
} The \ac{BLS} uses primary and complementary suppression to protect these tables \citep[p. 47]{spwp22}, but
the published summaries are exactly equal to the summaries from the confidential data. The suppressed cells can be reconstructed with great precision using the time series of county tables and known relationships among the cells. Since many of the suppressed cells contain just one or two business establishments, the reconstruction attack exposes those businesses to re-identification of their payroll and employment data.

Our last example is a database reconstruction attack that always produces exact re-identification.
Genome-wide association studies (GWAS) publish the marginal distributions of hundreds of thousands of alleles from study populations that are diagnosed with the same disease. It is possible to determine if a single genome was used to construct the GWAS with very high precision \citep{10.1371/journal.pgen.1000167,YU2014133,dwork:etal:2015}. An attacker who determines that a genome is in the GWAS learns that it is associated with the diagnosis defining the study population. This is a reconstruction attack because it establishes exact genomes that were input records for the GWAS. It is a re-identification attack because the attacker learns the medical case identifier associated with the genome. That association is unique due to human biology unless the person is an identical twin.\footnote{\citet{NOT-OD-14-124} revised the NIH rules for sharing genomic data, even when de-identified in a GWAS, to require active consent: ``[t]he final GDS [Genome Data Sharing] Policy permits unrestricted access to de-identified data, but only if participants have explicitly consented to sharing their data through unrestricted-access mechanisms.''}

\subsection{The Choice of an Accuracy Measure}
\label{sec:accuracy_intro}
We define statistical accuracy in terms of expected squared-error loss (Definition \ref{def:accuracy}).
One might hope for an analysis where the trade-off between privacy and accuracy is independent of how the statistics are used (for instance, independent of the prior beliefs of a Bayesian receiver).
In that case, for any two publications, all consumers would agree about the difference in published accuracy.
\citet{BrennerNissim:Impossibility:SIAM:2014} show that such universal mechanisms are impossible except in the special case of publishing a single counting query \citep{GhoshRoughgardenSunararajan:Universally:SIAM:2012}.
In Section \ref{sec:production}, we characterize several publication mechanisms that implicitly define a functional relationship between privacy loss and squared-error loss. We will also revisit the question of productive efficiency. Before doing so, we provide the formal, technical definitions of the concepts laid out in this section.

	\section{Model Preliminaries} 
	\label{sec:preliminaries}

We model a trusted custodian---the statistical agency---that controls a database from which it must publish population statistics.
We represent the database as a matrix with known properties, and the published statistics, called queries, as functions of the data matrix.
These queries represent statistics such as contingency tables and other standard public-use products.

\subsection{The Database}
The population database is a matrix, $D$.
Each row of $D$ contains information for one of $N$ individuals, and each column records a separate variable or feature.
$D$ is a multi-set with rows selected from a discrete, finite-valued \emph{data domain}, $\chi$.\footnote{In statistics, $\chi$ is called the sample space---the list of legal records in the database. Events that are deemed impossible a priori, structural zeros, can be accommodated in this framework. The assumption that $\chi$ is finite is not restrictive since, in practice, continuous data have discrete, bounded representations.}
We denote by $\left|\chi \right|$ the cardinality of $\chi$. This setup is very general.
It can handle missing data, non-response, skip patterns, alternative sources, unique identifiers, and other complex features of real data.

\subsubsection{Representation of the Database as a Histogram}
The \emph{histogram representation} of $D$ is a $\left|\chi\right| \times 1$ vector, $x \in  \mathbb{Z}^{\ast |\chi |}$,
where $\mathbb{Z}^{\ast}$ is the set of non-negative integers.
The histogram records the frequency of each feasible combination of attributes in $D$.
For each element of the data domain, $k\in\chi$, $x_{k}$ is the number of records in $D$ with attribute combination $k$.
The ordering of $k\in\chi$ is fixed and known, but arbitrary.

The $\ell _{1}$ norm of $x$ is
$\left\Vert x\right\Vert_{1}=\sum_{i=1}^{|\chi |}\left\vert x_{i}\right\vert=N$,
the number of records in the database.
Given two histograms ${x}$ and ${y}$, $\left\Vert{x}-{y}\right\Vert_{1}$ measures the number of
records that differ between ${x}$ and ${y}$.
\textit{Adjacent histograms} are those for which the $\ell _{1}$ distance is $1$.%
\footnote{%
If $x$ is the histogram representation of $D$, $y$ is the histogram
representation of $D^{\prime },$ and $D^{\prime }$ is constructed from $D $
by deleting or adding exactly one row, then $\left\Vert{x}-{y}\right\Vert_{1}=1$.
}

\subsubsection{Population Statistics Are Database Queries}
Population statistics are functions that map the data histogram to some output range $\mathcal{R}$.
A \emph{database query} is $q:\mathbb{Z}^{\ast |\chi |}\rightarrow \mathcal{R}$.
We call $q(x)$ the \emph{exact query answer}.

The \emph{sensitivity} of a query measures the maximum amount by which the exact answer can change when $D$ changes by the addition or removal of exactly one row.
The $\ell_{1}$ sensitivity for query $q$ is defined as
\begin{definition}[$\ell_{1}$ Query Sensitivity]
\label{def:query_sensitivity}
\begin{equation*}
	\Delta q = \max_{x,y\in \mathbb{Z}^{\ast |\chi |}, \left\Vert{x}-{y}\right\Vert_{1} \le 1}|q(x)-q(y)|.
\end{equation*}
\end{definition}
Sensitivity is a worst-case measure of how much information a given query can reveal. It is important in our analysis of privacy.

Most official statistical publications can be represented by linear queries.
A \textit{linear query} is $q(x) = q^{T}x$ where $q\in\left[-1,1\right]^{\left|\chi \right|}$.
A \textit{counting query} is a special case in which  $q \in \{0,1\}^{\left|\chi \right|}$.
Any marginal total from the fully-saturated contingency table representation of the database can be represented by a linear query.
The tables for legislative redistricting, for example, are among millions of marginal tables published from the decennial census.

The statistical agency wants to publish answers to a \emph{query workload}, $Q(\cdot)=\{q_{1}(\cdot),\ldots,q_{k}(\cdot) \}$.
The exact answer to the query workload on the histogram $x$ is a set $Q(x)=\left\{q_{1}(x),\ldots,q_{k}(x)\right\}$.
In the absence of privacy concerns, the statistical agency would publish $Q(x)$.
When the workload queries are linear, we represent the workload as a $k\times\left|\chi \right|$ matrix $Q$, which is the vertical concatenation of the $k$ scalar-valued linear queries. In this case, with some abuse of notation, we say $Qx$ is the exact answer to the query workload $Q(x)$.\footnote{It will be obvious from the context whether we are discussing the query workload, $Q(\cdot)$ or its matrix representation.} We extend Definition \ref{def:query_sensitivity} to this setting in Section \ref{sec:matrix_mechanism}.

\subsection{The Data Publication Mechanism}
As in computer science, we model the data publication mechanism as a stochastic function.\footnote{Deterministic mechanisms are implicitly included as a special case, i.e., with probability one.  Only trivial deterministic mechanisms are differentially private---``publish nothing'' or ``publish a constant.''
The distortion added to the exact query answer through the publication mechanism should enhance privacy, but will also reduce the accuracy of the published statistics.}

\begin{definition}[Data Publication Mechanism]
\label{def:query_mechanism} Let $\mathcal{F}$ be the set of allowable query workloads.
A \emph{data publication mechanism} is a random function
$M:\mathbb{Z}^{\ast |\chi|}\times \mathcal{F}\rightarrow \mathcal{R}$ whose
inputs are a histogram $x\in \mathbb{Z}^{\ast |\chi |}$ and a workload $Q \in  \mathcal{F}$,
and whose random output is an element of range $\mathcal{R}$.
For  $B \in \mathcal{B}$, where $\mathcal{B}$ are the measurable subsets of $\mathcal{R}$,
the conditional probability is $\Pr \left[ M(x,Q)\in B | x,Q \right] $, given $x$ and $Q$, where
the probabilities are only over the randomness induced by the mechanism.
\end{definition}

\subsubsection{Differential Privacy}
Our definition of differential privacy follows \citet{Dworketal:2006} and \citet{Dwork:Algorithmic:Book:2014}.
\begin{definition}[$\protect\varepsilon$-differential privacy]
	\label{def:dif_priv} Data publication mechanism $M$ satisfies $\varepsilon$%
	-dif\-fer\-en\-tial privacy if for all $\varepsilon > 0$, all $x,x^{\prime }\in
	N_{x}$, all $Q \in  \mathcal{F}$, and all $B\in \mathcal{B}$
	\begin{equation*}
	\Pr \left[ M(x,Q)\in B \ |x,Q\right] \leq e^{\varepsilon }\Pr \left[
	M(x^{\prime },Q)\in B \ |x^{\prime},Q\right],
	\end{equation*}%
	where $N_{x}=\left\{ (x,x^{\prime })\ s.t.~x,x^{\prime }\in \mathbb{Z}^{\ast |\chi |}~\text{and}~{||x-x^{\prime }\ ||_{1}=1}\right\} $ is the set of all
	\textit{adjacent histograms} of $x$, and as in Definition \ref{def:query_mechanism} the probabilities are taken only over the randomness in the mechanism.\footnote{Mechanisms satisfying Definition \ref{def:dif_priv} have several important properties. One that we use heavily is closure under composition, which means that if mechanism $M_1$ is $\varepsilon_1$-differentially private and mechanism $M_2$ is $\varepsilon_2$-differentially private, then the combination $M_{1,2}$ is $\varepsilon_1+\varepsilon_2$ differentially private. In our case, the composed mechanism is $Q_1, Q_2 \in \mathcal{F}$ and $M_{1,2} \equiv M(x,[Q_1,Q_2])$. For a proof in the general case see \citet[Chapter 3]{Dwork:Algorithmic:Book:2014}.}
\end{definition}

\subsubsection{Accuracy and Empirical Loss}
We define accuracy in terms of the squared $\ell_2$ distance between the mechanism output and the exact answer $Q(x)$.
\begin{definition}[Accuracy ($I$)]
  \label{def:accuracy}
  Given histogram $x \in \mathbb{Z}^{\ast |\chi |}$ and query workload $Q \in  \mathcal{F}$, the data publication mechanism $M(x,Q)$ has accuracy $I$ if
  $$\mathbb{E}\left[\left\Vert(M(x,Q) - Q(x)\right\Vert^{2}_{2}\right] = -I.$$ where $I\le 0$ and the expectation is taken only over the randomness in $M(x,Q)$.
\end{definition}
The notation $\left\Vert\cdot\right\Vert^{2}_{2}$ is the square of the $\ell_{2}$ (Euclidean) distance.
Accuracy is usually defined in terms of a positive expected loss, $\alpha=-I$, rather than in terms of the additive inverse. We use the additive inverse, $I$, so we can model accuracy as a ``good'' rather than a ``bad'' in what follows. We make this normalization for rhetorical convenience, and it is without mathematical consequence.\footnote{Readers familiar with the computer science literature may wonder why we model accuracy in terms of the expected loss rather than the worst-case accuracy used, for example, in \citet{Dwork:Algorithmic:Book:2014}.
Expected squared-error loss is more familiar to economists.
Our framework is readily extended to the other loss measures that appear in the differential privacy literature.
}



	\section{Differentially Private Publication as a Production Technology}
	\label{sec:production}

We explicitly characterize the production possibilities facing a statistical agency that publishes data using a known $\varepsilon$-differentially private mechanism.
Doing so allows the agency to make an explicit guarantee regarding protection against social privacy loss; that is, to make a verifiable claim about the value of the parameter controlling the non-rival public good parameterized by $\varepsilon$.
We describe two illustrative cases in which analysis of the mechanism yields a known technical relationship between accuracy and  privacy loss.
Furthermore, the function relating privacy loss and accuracy may be closed, bounded, and convex.
When all these properties hold, there exists a known marginal rate of transformation between privacy loss and accuracy.
It follows that the level of privacy loss chosen by the statistical agency entails an associated marginal cost of increasing privacy measured in units of foregone statistical accuracy.\footnote{Our analysis is not meant to be a descriptive, or positive, account of how statistical agencies or data custodians actually behave. It is explicitly normative.
}

\subsection{The Production of Privacy and Statistical Accuracy}
\label{sec:prod_basic}
Data publication mechanisms entail some bound on privacy loss ($\varepsilon$) and a level of statistical accuracy ($I$).
Any mechanism is associated with a pair $(\varepsilon,I)$.
Following standard producer theory, we refer to each pair as a \emph{production activity}.
Production activities usually represent vectors of inputs and outputs such that the inputs can be transformed into the outputs.
Our insight is to think of the information in the database as akin to an endowment of potential privacy loss.
Some of the privacy loss endowment must be expended by the data custodian to produce population statistics of any accuracy.

The \emph{transformation set}, $Y$, contains all feasible production activities available to the statistical agency. We assume $Y$ is closed. We also assume inactivity is possible, but obtaining non-trivial accuracy requires some privacy loss (no free lunch). Likewise, obtaining perfect accuracy ($I=0$) requires infinite privacy loss ($\varepsilon=\infty$).
Under these assumptions, we can represent $Y$ with a \emph{transformation function} $G(\varepsilon,I)$ such that
$
Y=\left\{ \left( \varepsilon ,I \right) \left\vert \varepsilon >0,I<0 \text{ s.t. }G(\varepsilon ,I)\leq0
\right\}\right\}.
\label{eqn:transformation_set}
$
The \emph{production frontier} is the set
\begin{equation}
PF=\left\{ \left( \varepsilon ,I\right)
\left\vert \varepsilon >0,I<0\text{ s.t. }G(\varepsilon ,I)=0\right.
\right\} .  \label{eqn:ppf}
\end{equation}%
Equation (\ref{eqn:ppf}) yields an implicit functional relationship between $\varepsilon$ and $I$. As a theoretical proposition, (\ref{eqn:ppf}) provides guidance on the constructs at the heart of our analysis. Its direct implementation is problematic. Given the current state of knowledge, discussed explicitly in Section \ref{sec:efficiency}, there is no general solution for $G(\varepsilon,I)$. The statistical agency must select the best available technology for the query workload of interest and implement the $G$ that technology implies. The agency should be guided by knowledge of recent advances and known impossibility results, but it cannot yet rely on algorithms known to solve equation (\ref{eqn:ppf}) for general query workloads.

Scarcity is a key feature of the economic theory of production, often expressed in the axiom of ``no free lunch.'' As it turns out, there is no free lunch when it comes to data privacy. In separate contributions, \citet{Dwork2004}, \citet{Dwork2008}, \citet{gehrke2011towards}, and \citet{Kifer:2011:NFL:1989323.1989345} all show that publishing useful statistical summaries requires an explicit loss of privacy. This result holds for any non-trivial definition of formal privacy including, but not restricted to, differential privacy.

\subsection{Example: Randomized Response}
\label{sec:randomized_response}
To build intuition, we offer a brief, but accessible, example of a differentially private data publication mechanism known as \emph{randomized response} \citep{Warner:RandResponse:JASA:1965}. The statistic of interest is the population proportion of a sensitive binary char\-ac\-ter\-is\-tic---for example, whether the respondent voted for a particular candidate. Randomized response protects the individual against privacy loss even when his ``yes'' or  ``no'' response can be attributed directly to him. It does so by creating uncertainty about whether the respondent answered the sensitive question or some non-sensitive question.

In a survey setting, the respondent draws a sealed envelope. In it, the respondent finds one of two yes/no questions. The interviewer records the response, but does not know, and cannot record, which question was answered.\footnote{The sensitive data can also be collected automatically by a web browser, which performs the randomization in the background before transmitting information to an analyst. This approach is formalized in a tool known as \ac{RAPPOR}, and used by Google's Chrome browser to detect security problems \citep{Erlingsson2014,DBLP:journals/corr/FantiPE15}.}
The data analyst knows only that the envelope contained the sensitive question with probability $\rho$ and an innocuous question with probability $1-\rho$.

Randomized response guarantees privacy to everyone. Those who answer the sensitive question can be assured that no one, including the interviewer, knows their response with certainty. We can measure the amount of information leaked about the sensitive characteristic. Finally, privacy increases with the probability that the innocuous question is asked. However, this also increases the uncertainty about the distance between the published statistic and the population proportion of the sensitive characteristic.

To formalize randomized response in terms of the model of this paper,
suppose the population database consists of two vectors $x$ and $z$.
Each entry $x_{i}$ is a binary indicator corresponding to the sensitive trait for $i$.
Each entry $z_{i}$ corresponds to the non-sensitive trait.
The statistical agency publishes a vector $d$ that is conformable with $x$ by randomizing over whether it reports the sensitive or the non-sensitive characteristic.
Let $T_{i}$ be a Bernoulli random variable that determines which entry is reported.
The published value $d_{i}=T_{i}x_{i} + (1-T_{i})z_{i}$.
To complete the description, denote $\rho=Pr\left[T_{i}=1\right]$ and $\mu=Pr\left[z_{i}=1\right]$.
To make privacy guarantees, we require $0<\rho < 1$.
We also assume that $z_{i}$ is independent of $x_{i}$.
For convenience, we assume $\mu=0.5$.\footnote{The justification for, and implications of, this assumption are elaborated in Appendix \ref{app:randomized_response}. This assumption is without consequence for our illustration.}

Since the randomization is independent for each entry $i$, we can restrict attention to the conditional probabilities
\begin{align}
	Pr\left[d_{i}=1|x_{i}=1\right] &= \rho+0.5(1-\rho) \\
	Pr\left[d_{i}=1|x_{i}=0\right] &= 0.5(1-\rho).
\end{align}
Differential privacy bounds the ratio of these two probabilities as well as the ratio associated with the event $d_{i}=0$.
Randomized response is $\varepsilon$-differentially private with
\begin{equation}
 	\varepsilon(\rho) =\log\left(\frac{1+\rho}{1-\rho} \right).
 	\label{eq:rand_response_epsilon_func}
 \end{equation}
Semantic, or inferential, privacy concerns what can be learned about the sensitive characteristic conditional on what is published, $Pr\left[x_{i}=1|d_{i}\right]$. By Bayes' rule, the bound in \eqref{eq:rand_response_epsilon_func} also applies to posterior inferences about the sensitive characteristic (see Appendix \ref{app:randomized_response}).

The goal of the analysis is to draw inferences about the population proportion of the sensitive characteristic.
Define $\widehat{\beta} = \frac{1}{N}\sum_{i}d_{i}$, the empirical mean proportion of ones in the published responses, and $\pi = \frac{1}{N}\sum_{i}x_{i}$, the (unobserved) mean of the sensitive characteristic.
Finally, define $\widehat{\pi}\left( \rho\right) = \frac{\widehat{\beta}-\mu(1-\rho)}{\rho}$, which is an unbiased estimator of $\pi$ with variance $\limfunc{Var}[\widehat{\pi}\left( \rho\right)] = \frac{\limfunc{Var}[\widehat{\beta}]}{\rho^{2}}$.
Therefore, accuracy is:
$
 	I(\rho) =  - \limfunc{Var}[\widehat{\pi}\left( \rho\right)].
$
In Appendix \ref{app:randomized_response} we also show $\frac{dI}{d\varepsilon}>0$ and $\frac{d^{2}I}{d\varepsilon^{2}}<0$.
Therefore, the technical relationship between privacy-loss ($\varepsilon$) and accuracy ($I$) is strictly increasing and concave.

\subsection{Example: Matrix Mechanism}
For the rest of the paper, we consider a statistical agency that publishes statistics using the \emph{matrix mechanism} introduced by \citet*{li:matrix:VLDB:2015}.
Unlike randomized response, which operates directly on the micro-data record, the matrix mechanism is a general class of data-independent mechanisms that protect privacy by adding noise to the exact query answers that is calibrated to the query workload sensitivity. The matrix mechanism is also under active development for use by the Census Bureau \citep{mckenna2018optimizing,Kuo:DPHierarchical:CoRR:2018}.
Our analysis is generally valid for all differentially private mechanisms that yield a convex relationship between privacy loss and accuracy.\footnote{%
In previous versions of this paper, we have considered data-dependent mechanisms, including the Multiplicative Weights Exponential Mechanism (MWEM) introduced by \citet*{Hardt:Simple:NIPS:2012} and the Private Multiplicative Weights (PMW) mechanism, due to \citet*{Hardt:Multiplicative:FOCS10}. While these mechanisms also yield well-behaved relationships between privacy and accuracy, their dependence on the data means their accuracy guarantees can only be stated in terms of the worst-case absolute error across all queries, not the expected square-error accuracy measure we focus on in this paper. This reinforces our observation that the definition of accuracy is not without loss of generality.
}

\subsubsection{The Matrix Mechanism}
\label{sec:matrix_mechanism}
To introduce the matrix mechanism, we first describe the simpler Laplace mechanism operating on a query workload.
\citet{Dworketal:2006} proved the single query version of the Laplace mechanism, which \citet{li:matrix:VLDB:2015} generalized to a matrix workload. We state the matrix version with new notation: $\Delta Q$ in Theorem \ref{prop:laplace_mechanism} is the generalization of $\ell_{1}$-sensitivity for the query workload (defined formally in Appendix \ref{app:matrix_mechanism}).
\begin{theorem}[Laplace Mechanism]
\label{prop:laplace_mechanism}
 For $\varepsilon>0$, query workload $Q$, and histogram $x$, define data publication mechanism
 $\limfunc{Lap}(x,Q) \equiv Qx + e$,
 where $e$ is a conformable vector of iid samples drawn from the Laplace distribution with scale parameter
 $b=\frac{\Delta Q }{\varepsilon}$.
 $Lap(x,Q)$ is $\varepsilon$-differentially private.
\end{theorem}
For the proof, see \citet[prop. 2]{li:matrix:VLDB:2015}.  The amount of noise added by the Laplace mechanism increases with the workload query sensitivity $\Delta Q$ and decreases with $\varepsilon$.\footnote{A Laplace random variable---called ``double exponential'' by statisticians---has density  $(1/2b) \exp(-|x|/b)$ on the real line with $\mathbb{E}\left[e\right]=0$ and $\limfunc{Var}\left[e \right]=2b^{2}$. It is more peaked and has fatter tails than the normal distribution \citep[p. 31]{Dwork:Algorithmic:Book:2014}. Its discrete equivalent is called the geometric distribution, and the associated geometric mechanism is also differentially private \citep[p. 1674-5]{GhoshRoughgardenSunararajan:Universally:SIAM:2012}.}

The matrix mechanism improves on the Laplace mechanism by finding a query strategy matrix $A$, conformable with $Q$, such that each query in $Q$ can be expressed as a linear combination of queries in $A$. The idea is to find a strategy matrix $A$ with lower sensitivity than $Q$, thereby allowing greater accuracy for any particular level of privacy loss.
\begin{theorem}[Matrix Mechanism Implemented with Laplace Mechanism]
\label{thm:ppf} For histogram $x$, query workload $Q$, query strategy $A$, and $\varepsilon >0$, the matrix mechanism $M(x,Q)$ publishes $Qx + QA^{+}\left(\Delta A\right)e$, where $e$ is a vector of iid Laplace random variables with mean zero and scale parameter $b=1 / \varepsilon$.

\begin{enumerate}
\item \emph{Privacy}: The matrix mechanism is $\varepsilon$-differentially private.

\item \emph{Accuracy}: The matrix mechanism has accuracy $
I = - \limfunc{Var}(e) \left(\Delta A\right)^{2}\left\Vert QA^{+}\right\Vert_{F}^{2},
$
where $A^{+}$ is the Moore-Penrose inverse of $A$, $\Delta A$ is the generalization of $\ell_{1}$-sensitivity for the query workload (see Appendix \ref{app:matrix_mechanism}). The notation $\left\Vert \cdot \right\Vert_{F}$ refers to the matrix Frobenius norm, which is the square root of the sum of the absolute squared value of all elements in the vector or matrix between the braces \citep[p.~55]{GolubVanLoan:MatrixComputations:1996}.
\end{enumerate}

\end{theorem}
For the proof, see \citet[prop. 7]{li:matrix:VLDB:2015}.\footnote{The strategy matrix $A$ is not hypothetical. \citet[p.~768-78]{li:matrix:VLDB:2015} provide examples with and without side constraints of algorithms that successfully choose $A$. \citet*{mckenna2018optimizing} demonstrate the feasibility of computing $A$ for large query workloads like those found in decennial censuses by exploiting Kronecker products. Nevertheless, we acknowledge that for general, large, linear query workloads, the computation of a solution for $A$ is an unsolved problem.}

When the queries in $A$ are answered using the Laplace mechanism, the implied marginal cost of increasing accuracy $I$ in terms of foregone privacy protection $\varepsilon$---the marginal rate of transformation---is%
\begin{equation}
MRT \equiv \frac{dI}{d\varepsilon }=-\frac{%
\partial G/\partial \varepsilon }{\partial G/\partial I}= \frac{4 \left(\Delta A\right)^{2}\left\Vert QA^{+}\right\Vert_{F}^{2}}{\varepsilon ^{3}}.  \label{eqn:MRT}
\end{equation}%
The marginal rate of transformation is positive because privacy loss is a public bad.\footnote{The $MRT$ is not hypothetical. The accuracy guarantee in Definition (\ref{def:accuracy}) is exact. The production function is exact in the two public goods. The requirements $dI/d\varepsilon >0$ and $d^{2}I/d^{2}\varepsilon < 0$ can be verified by substituting $\limfunc{Var}(e) = 2/\varepsilon^{2}$ and differentiating. The matrix mechanism can be operated with any data-independent mechanism, such as the geometric mechanism, replacing the Laplace mechanism in its definition.}

\subsection{Efficiency in Production}
\label{sec:efficiency}
We would like to know whether the matrix mechanism provides maximal accuracy for any choice of $\varepsilon$, or if it is possible to achieve greater accuracy.
For mechanisms involving multiple linear queries, \citet{Hardt:OnGeometry:STOC:2010} established upper and lower bounds on the variance of the noise added to achieve privacy. Their lower bound result implies there is a maximal attainable level of accuracy for any mechanism providing $\varepsilon$-differential privacy.
They also established the existence of a differentially private mechanism, the \emph{K-norm mechanism}, that approximately achieves maximal accuracy.\footnote{Their results were subsequently refined and extended by \citet{Nikolov:Geometry:STOC:2013} and \citet{Bhaskara:2012:UDP:2213977.2214089}.}
In principle, our matrix mechanism could be operated with the K-norm mechanism instead of the Laplace mechanism.
However, there is no closed-form solution for accuracy when using the K-norm mechanism, and it is computationally burdensome.\footnote{Computer scientists acknowledge the practical need to trade off computational costs against privacy and accuracy \citep[p.~50]{Vadhan:Complexity:2017}. We use the Laplace mechanism instead of the K-norm mechanism among data independent mechanisms for this reason.
}

	\section{Preferences for Privacy and Social Choice}
	\label{sec:preferences}
In this section we formally relate the definition of differential privacy to a semantic interpretation of privacy as the protection of individual secrets. We then develop a basic model of preferences for privacy and accuracy. We also derive the formal statement that the optimal levels of privacy protection and accuracy are determined by setting the marginal rate of transformation equal to the marginal willingness to accept privacy loss.

\subsection{Differential Privacy as a Bound on Learning}
\label{sec:dp_semantics}
Following \citet{Kifer2012}, we describe necessary and sufficient conditions under which differential privacy implies a bound
on what an attacker can learn about any person's sensitive data from published statistics. We call this the \emph{semantic privacy} bound.
We call the sensitive data items \emph{secrets}.
A secret pair consists of two mutually exclusive events---$s_{i}$ and $s_{i}^{\prime}$.
The event $s_{i}$ means the record for individual $i$ was included in the database and has attribute $\chi_{a}$.
The event $s_{i}^{\prime}$ means the data for $i$ was not included in the database.

Suppose the statistical agency publishes statistics using a $\varepsilon$-differentially private mechanism.
Semantic privacy is the change in the odds of $s_{i}$ versus $s_{i}^{\prime}$ for an attacker before and after the statistics are published.
 Inference requires a model of the random process that generates the database, $Pr\left[D\ |\ \theta\right]$. It is characterized by a parameter $\theta$, and that the sampling probabilities for all individuals are independent of one another.\footnote{See Appendix \ref{app:dp_semantics_details} for details.}
Applying Bayes' law to Definition \ref{def:dif_priv} and using the data generating process in Appendix equation \eqref{eq:dgp}, differential privacy implies
\begin{equation}
   e^{-\varepsilon} \le \frac{Pr[s_i\ |B,\theta]}{Pr[s_{i}^\prime\ |B,\theta]} \Biggm/
   \frac{Pr[s_i\ |\theta]}{Pr[s_{i}^\prime\ |\theta]} \le e^{\varepsilon}.
   \label{eq:semanticbound}
\end{equation}
For the proof, see Theorem 6.1 in \citet{Kifer2012}.\footnote{Privacy semantics were defined as $\varepsilon$-in\-dis\-tin\-guish\-abil\-ity, by \citet{dwork:2006}, one of the original differential privacy papers. This inference system is not Bayesian learning, and for our case, generates a semantic bound that is also $[-\varepsilon,\varepsilon]$ even though it is defined over arbitrary data generating processes. \citet{kasiviswanathan2014semantics} also generate a semantic bound that can be computed from the differential privacy bound without assumptions on the data generating process. \citet{Wasserman:Statistical:JASA:2010} have given an interpretation of differential privacy as bounding the power of a statistical test of the null hypothesis about the value of a secret. \citet[Section 4.3]{Nissim:DPNonTech:WP:2018} provides an extremely lucid non-technical description of privacy semantics in terms of posterior-to-posterior comparisons.}

Equation \eqref{eq:semanticbound} says that when the agency publishes using an $\varepsilon$-differentially private mechanism,
the Bayes factor associated with any secret pair is bounded by the same $\varepsilon$.
The semantics depend on the data generating process, and the conditioning on $\theta$ is nontrivial. As Kifer and Machanvajjhala show, the data generating process in equation (\ref{eq:dgp}) is the only one for which mechanisms satisfying differential privacy generate semantics satisfying equation (\ref{eq:semanticbound}).

Being able to move between the bounds on the mechanism and the bounds on the privacy semantics
is critical to applying differential privacy to the social choice problem in this paper. When we derive technology sets, these bounds constrain the feasible pairs $(\varepsilon,I)$; hence, the $\varepsilon$ bounds on the mechanism implied by the definition of differential privacy (Definition \ref{def:dif_priv}) are important. When we model preferences for $(\varepsilon,I)$, individuals care about the mechanism's ability to protect secrets learned from the published statistics; hence, the $\varepsilon$ semantic bounds in equation (\ref{eq:semanticbound}) matter.

\subsection{Modeling Preferences over Privacy and Accuracy}
\label{sec:prefs_main}
We assume the statistical agency chooses $\varepsilon$ and $I$ to maximize a utilitarian social welfare
function%
\begin{equation}
SWF\left( \varepsilon ,I\right) =\sum_{i}v_{i}\left(\varepsilon ,I\right)  \label{eqn:swf}
\end{equation}%
subject to the restriction that it operates along the production frontier \eqref{eqn:ppf}.
The functions $v_{i}\left(\varepsilon ,I\right) $ measure indirect utility for each person $i$.
Indirect utility depends on the levels of $\varepsilon$ and $I$.
Our formulation allows for arbitrary heterogeneity in preferences for both privacy loss and statistical accuracy.
In doing so, we allow for the standard case in which one group cares only about privacy, while another group cares only about accuracy.
Following \citet{Nissim:2012:PMD:2229012.2229073}, we assume utility is additively separable into information utility and data utility: $v_{i}\left(
\varepsilon ,I\right) = v^{Info}_{i}(\varepsilon) + v^{Data}_{i}(I)$.\footnote{There is no obvious consumption complementarity between privacy and accuracy in the setup we consider. This assumption could be violated if, say, accuracy indirectly affects utility by increasing wealth or by decreasing the prices of physical goods.}

\subsubsection{Information Utility}
\label{subsec:info_utility}
First, we specify individual information utility---preferences for privacy.
Our approach is motivated by \citet{Ghosh:Auction:GEB:2015}. Suppose that $\Omega$ is the set of future events, or states of the world, over which an individual has preferences.
These are states of the world that may become more likely if confidential information in the database is disclosed. This might correspond to identity theft, the threat of being persecuted for a particular trait, or denial of a health insurance claim.

Let individual $i$'s utility from event $\omega\in\Omega$ be $u_{i}(\omega)$. The individual's data may be used in an $\varepsilon$-differentially private mechanism $M(x,Q)$ with output drawn from range $\mathcal{R}$ according to the distribution induced by the mechanism. Finally, let $z(M(x,Q))$ be an arbitrary function that maps the published outcome of $M$ onto a probability distribution over events in $\Omega$. As discussed in Section \ref{sec:measuring_privacy}, differential privacy is invariant under post-processing. It follows that the transformation $z(M(x,Q))$ is also $\varepsilon$-differentially private because $z$ only uses outputs of $M$.

From the individual's perspective, what matters is the difference between what can be learned about their secrets when their information is included in the data, $x$, and when it is not. Let $x^\prime$ denote the neighboring histogram that excludes $i$'s data.
By differential privacy, $i$'s expected utility satisfies
$\mathbb{E}_{\omega|M(x,Q)}\left[u_{i}(\omega)\right]$ $\le$ $e^{\varepsilon}\mathbb{E}_{\omega|M(x^\prime,Q)}\left[u_{i}(\omega)\right]$.
The worst-case incremental utility loss from having their data included in the mechanism, is $\left(e^{\varepsilon}-1\right)v_{i}$ where $v_{i}$ is the expected utility over future events when $i$'s data are not included in the mechanism.
This argument supports a model of preferences in which $\varepsilon$ enters linearly.\footnote{When $\varepsilon$ is small, $\varepsilon\approx e^{\varepsilon}-1$. \citet{Ghosh:Auction:GEB:2015} consider a broader class of information utility models, of which the linear model as a special case}
 Following \citet{Ghosh:Auction:GEB:2015}, we assume $v^{Info}_{i}(\varepsilon) = -k_{i}\varepsilon$ where $k_{i}\ge 0$ to reflect that privacy loss measured by $\varepsilon$ is a public ``bad.''\footnote{\citet{Nissim:2012:PMD:2229012.2229073} observe that the framework in \citet{Ghosh:Auction:GEB:2015} is an upper bound and that expected utility loss may be lower. They show that knowing the upper bound is sufficient for certain problems in mechanism design. The presence of $\varepsilon$ in the utility function could also reflect the consumer's recognition that the statistical agency applies the same value of $\varepsilon$ to everyone. We would assume they have preferences over this property of the mechanism. In this case, the connection to expected harm from breaches of their secrets would not be as direct.}

\subsubsection{Data Utility}
\label{subsec:data_utility}

Next, we introduce a model that supports a reduced-form specification for data utility that is linear in accuracy: $v^{data}_{i}(I) =  a_{i} + b_{i}I$.
Our simple, but illustrative, model is a proof-of-concept. We hope to inspire further investigation of preferences for public statistics, since there is very little existing research on which to draw.\footnote{\citet{Spencer:Optimal:JASA:1985,Spencer:Needed:JASA:1990} and \citet{Seeskin:Spencer:Effects:2015} attempt to measure the social cost of inaccuracy of official statistics. The statistics literature generally defines data ``utility'' via a loss function. See, e.g., \citet{Trottini:Modelling:IJUFK:2002}.}

We associate the data utility for any person with her expected utility of wealth given her knowledge of the publication system.
To model this expectation, we assume wealth depends on the statistics published by the statistical agency, and that individuals are aware that error is introduced by the privacy protection system.
More concretely, each person $i$ gets utility from wealth according to a twice-differentiable and strictly concave function, $U_{i}(W_{i})$.
We let $W_{i} = \Pi_{i}^{T} M(x,Q)$, where $M(x,Q)$ is the vector of published population statistics and $\Pi_{i}$ is a person-specific vector of weights.
We require only that the entries of $\Pi_{i}$ be finite.
As described in Theorem \ref{thm:ppf}, $M(x,Q) = Qx + Q(\Delta A)A^{+}e$ where $A$ is a query strategy matrix with pseudo-inverse $A^{+}$ and $e$ is a vector of $iid$ random variables with $\mathbb{E}\left[e\right]=0$ and whose distribution is independent of $x$, $Q$, and $A$.

The query set $Q$ and the distribution of $e$ are public knowledge---a key feature of differential privacy. Hence, uncertainty is with respect to the data histogram $x$ and the realized disturbances $e$. Beliefs about the distribution of $x$ may be arbitrary and person-specific. The expected utility for any person $i$ is
$$
	\mathbb{E}\left[U_{i}(W_{i})\right] = \mathbb{E}_{x}\left[\mathbb{E}_{e|x}\left[U_{i}\left(\Pi_{i}^{T}Qx + \Pi_{i}^{T}QA^{+}(\Delta A)e\right)\vert x\right] \right].
$$
In Appendx \ref{app:data_utility_details}, we show this can be approximated as
 \begin{equation}
 	\begin{array}{rcl}
 	 \mathbb{E}\left[U_{i}(W_{i})\right]  \approx &  \mathbb{E}_{x}\left[U_{i}(\Pi_{i}^{T}Qx)\right] &- I\cdot \left\{\frac{1}{2}\mathbb{E}_{x}\left[U_{i}^{\prime\prime}(\Pi_{i}^{T}Qx)\right] \frac{\left\Vert \Pi_{i}^{T}QA^{+} \right\Vert^{2}_{F}}{\left\Vert QA^{+} \right\Vert^{2}_{F}}\right\} \\
 	 \equiv & a_{i} & + I\cdot b_{i}.
 	 \end{array}
 \end{equation}
 We obtain this result by taking expectations of a second-order Taylor series approximation to $i$'s utility around $e=0$.
 Our derivations and the result that expected utility is decreasing with the variance of wealth are familiar from the literature on risk aversion \citep[see ch.~1]{Eeckhoudt:EFDU:Princeton:2005}.
 The final expression highlights that from the planner's perspective, all that matters are person-specific weights associated with the utility of data accuracy.
We proceed with a reduced-form model for data utility that is linear in accuracy: $v^{data}_{i}(I) = a_{i} + b_{i}I$.

\subsubsection{Equilibrium}
Assuming the indirect utility functions are differentiable, the conditions that characterize the welfare-maximizing levels of
$\varepsilon$ and $I$ subject to the feasibility constraint are%
\begin{equation}
\frac{\frac{\partial G\left( \varepsilon ^{0},I^{0}\right) }{\partial
\varepsilon }}{\frac{\partial G\left( \varepsilon ^{0},I^{0}\right) }{%
\partial I}}=
	\frac{\frac{\partial }{\partial \varepsilon }%
\sum_{i}v^{Info}_{i}\left(\varepsilon ^{0}\right) }{\frac{%
\partial }{\partial I}\sum_{i}v^{Data}_{i}\left(I ^{0}\right)} = \frac{\sum_{i}k_{i}}{\sum_{i}b_{i}} = \frac{\bar{k}}{\bar{b}}.
  \label{eqn:optimum}
\end{equation}%
The left-hand side of
equation (\ref{eqn:optimum}) is the marginal rate of transformation based on the
production frontier while the right-hand side is the marginal rate of substitution between privacy loss and accuracy. We also refer to this as the \emph{willingness to accept} privacy loss measured in units of statistical accuracy.

To choose privacy and accuracy, an agency needs to know the marginal rate of transformation and the willingness to accept as shown in \eqref{eqn:optimum}. It must solve for the efficient query strategy $A$.\footnote{The choice of query strategy $A$ depends on the query workload, $Q$, the statistics of interest from the data collection, and the differentially private mechanism that will be used to answer $A$. See \citet{li:matrix:VLDB:2015} for details.  } Once it does, it knows the marginal rate of transformation at any point. The choice then depends on two unknown quantities: the average preference for privacy in the population ($\bar{k}$) and the average preference for accuracy ($\bar{b}$). In Section \ref{sec:titleI}, we calibrate these quantities in an analysis that illustrates the strengths and drawbacks of our approach.

\subsection{Accuracy and Privacy Protection Are Public Goods}
\label{sec:public_goods}
In our model, $\varepsilon$ and $I$ are public goods.
Once the statistical agency sets these parameters, all individuals enjoy the same statistics and privacy protection. However, our framework allows each person to have different preferences for privacy loss and accuracy.\footnote{If statistical accuracy is an input to production, consumer utility also depends on accuracy indirectly through prices. Note, too, that our approach is distinct from much of the economics of privacy research that considers how companies elicit information about consumers' idiosyncratic preferences and thereby engage in price discrimination \citep{Odlyzko:Privacy:EconInfoSecurity:2004}.}
It is natural to treat accuracy of official statistical publications as being both non-rival and non-excludable in consumption \citep{acquisti:taylor:wagman:2015}.
\citet{Hsu:EconomicEpsilon:IEEE:2014} and \citet{Ghosh:Auction:GEB:2015}  model the provision of statistical summaries for use by a private analyst.
Neither paper acknowledges the public-good nature of either the published statistics or the privacy protection afforded by the database custodian.

That privacy protection is a public good was foreshadowed by \citet[][p. 3]{dwork2008differential} when she wrote:
``\lbrack t]he parameter $\varepsilon $ ... is public. The choice of $\varepsilon $ is essentially a social question.''
This observation has not previously made its way into models of the market for privacy rights.
All persons receive the same guarantee against privacy loss, $\varepsilon$. That is, all persons participate in a data collection and publication system wherein the privacy-loss parameter is set independent of any characteristics of the population actually measured.
This is our interpretation of the \textquotedblleft equal protection under the
law\textquotedblright\ confidentiality-protection constraint that governs most national statistical agencies.
In addition, the benefits from increased privacy protection for any individual in the population are automatically enjoyed by every other individual, whether that person's information is used or not---the privacy protection is therefore strictly non-rival.

The public scrutiny of government statistical agencies create strong incentives to emphasize privacy protection \citep[Chapter 5]{groves:harris-kojetin:2017}. In practice, an agency's choice is governed by legal, economic, and political considerations. One might reasonably ask ``Why should the privacy bound $\varepsilon$ arrived at by a statistical agency for reasons related to policy or legislation be the quantification of privacy loss relevant to economic actors? In environments where privacy loss is largely determined by a much smaller population of uniques or of people particularly susceptible to re-identification, features of that subpopulation might be all that determines the privacy-accuracy production function.''\footnote{We are grateful to an internal reviewer at the Census Bureau for providing this argument.} This argument is flawed by its failure to acknowledge that in all statistical databases of which we are aware every single row is unique.\footnote{For instance, in databases of households, the household address is geocoded to its GPS coordinates. In databases of businesses, detailed geography and industry codes uniquely distinguish firms.}

When all data are unique, deciding how to publish meaningful statistics involves choices that compromise the privacy of some sub-populations more than others.
Conventional \ac{SDL} attempts to present summaries with granularity in one dimension (say detailed geography) and similar summaries in another dimension (say detailed racial categories) without having to account for the risk in the cross-classification (detailed geography by detailed race categories). The database reconstruction theorem exposes the vulnerability in that practice. If every record is really a population unique, then publication mechanisms that are closed under composition are necessary to keep track of the exposure from the cumulative set of published statistics. Such mechanisms require worst-case analysis. And this worst-case analysis means privacy protection is a non-rival public good.

	\section{Example: Title 1 School Funding Allocations} 
	\label{sec:title_I}
\label{sec:titleI}
Our first application is the allocation of federal funding to school districts under Title I.
If statistics on the number of Title I-eligible students in each school district are published using a differentially private mechanism, then funding will be misallocated due to the error induced by the mechanism.
However, publishing those statistics without protection compromises privacy.
We show how this social choice problem could be managed in practice.

\subsection{Setting}
Title I of the Elementary and Secondary Education Act of 1965 provides federal funding to help improve educational outcomes for disadvantaged students.
Funds are appropriated by Congress and then allocated to school districts based on need.
That need is determined, in part, using statistics published by the Census Bureau.

\citet{Sonnenberg:TitleI:NCES:2016} describes how Title I allocations are determined.
The \ac{DOE} allocates basic grants using a formula that depends on
$
	A_{\ell} = E_{\ell}\times C_{\ell},
$
where $A_{\ell}$ is the \emph{authorization amount} for school district $\ell$, $E_{\ell}$ is the \emph{eligibility count}, and $C_{\ell}$ is the \emph{adjusted \ac{SPPE}}.
To keep the analysis focused on the core privacy-accuracy trade-off, we assume this formula determines total funding to district $\ell$ and that $C_{\ell}$ is known with certainty, but the DOE must use a published count of Title I-eligible students $\widehat{E}_{\ell}$ that may differ from the true count due to privacy protection.\footnote{To be very clear, this does not describe the actual data collection or statistical disclosure practices currently used by the Census Bureau.
Our example also abstracts from other types of non-survey error. These considerations are important, as \citet{Manski2014} correctly argues. Our analysis shows that errors from privacy protection can be meaningful, even in the absence of other concerns about data quality.}

The DOE is supposed to allocate $X = \sum_{\ell=1}^{L} E_{\ell}\times C_{\ell}$ dollars under Title I, but the actual allocation will be $\widehat{X} = \sum_{\ell=1}^{L} \widehat{E}_{\ell}\times C_{\ell}$.
The policy challenge is to balance privacy loss among disadvantaged households against the misallocation of Title I funds.

\subsection{The Social Choice Problem}
We assume the Census Bureau has data on the complete population of school age children that includes two fields: the school district and an indicator for whether the child counts toward Title I eligibility. It publishes privacy-protected counts of the total number of Title I-eligible children in each district using the matrix mechanism of Theorem \ref{thm:ppf}. The mechanism returns $\widehat{E}_{\ell} = E_{\ell}+e_{\ell}$ where $e_{\ell}$ is Laplace noise. By Theorem \ref{thm:ppf}, the published counts satisfy $\varepsilon$-differential privacy.\footnote{Because school districts do not overlap, the query workload sensitivity for the entire country is the same as the sensitivity for a single district---namely $1$. The guarantee of $\varepsilon$-differential privacy is the same for each district separately---a property called \emph{parallel composition}.
The privacy loss for a student in one school district is not affected by what is guaranteed to students in other districts. This does not mean that learning the number of Title I students in district $A$ is uninformative about Title I status of students in district $B$. If the districts are neighboring, the statistical outcomes may be correlated, and differentially private mechanisms permit learning about this correlation.
}
We also know the accuracy is:
\begin{equation}
	I = -\mathbb{E}\left[\sum_{\ell=1}^{L}\left(\widehat{E}_{\ell}-E_{\ell}\right)^{2}\right] = -\frac{2L}{\varepsilon^{2}}
	\label{eq:prod_function_TitleI}
\end{equation}
where $L$ is the total number of school districts.\footnote{Adding Laplace noise can result in published counts that are non-integer and potentially negative. Edits to enforce range restrictions---either by the Census Bureau prior to publication, or by the Department of Education prior to use---have only minor consequences for this analysis. The postprocessing does not affect the privacy guarantee, but it can affect the accuracy. See \citet{Li2012}.
}

Suppose policymakers' choices are guided by a variant of the social welfare function studied in Section \ref{sec:preferences}.
Specifically:
\begin{equation}
	SWF = \phi\sum_{i}v_{i}^{Info}(\varepsilon) + (1-\phi)v^{Data}(I),
\end{equation}
where the first summand on the right-hand-side reflects the linear aggregation of individual utility from privacy loss, and the second summand reflects the social cost of misallocating funds. The parameter $\phi$, which is new, is the weight on privacy in total welfare.


As in Section  \ref{subsec:info_utility}, let $v_{i}^{Info}(\varepsilon) = -k_{i}\varepsilon$ reflect the incremental loss to utility for person $i$ associated with having her data used in a publication with privacy guarantee $\varepsilon$.
Regarding $v^{Data}(I)$, the social planner's preferences are linear-quadratic in the aggregate misallocation $W=(\widehat{X}-X) = \sum_{\ell=1}^{L}C_{\ell}\left[\widehat{E}_{\ell}-E_{\ell}\right]$ so that
$
	v^{Data}(I) = I\sum_{\ell=1}^{L}\frac{C_{\ell}^{2}}{L}
$.\footnote{Here, we assume the planner has a distaste for misallocation. We could instead adopt a more general model for data utility following the analysis in Section \ref{subsec:data_utility} that would associate the planner's preferences for accuracy with the expected utility of each district. Doing so would add another set of utility parameters to model and calibrate and distract from our primary goal of offering a simple, stylized illustration of the approach developed in this paper.
}
Following \eqref{eqn:optimum}, the social planner's willingness to accept privacy loss is
\begin{equation}
	WTA \equiv \frac{dI}{d\varepsilon} = \eta\frac{\sum_{i=1}^{N}k_{i}}{\overline{C^{2}}} = \eta N\frac{\bar{k}}{\overline{C^{2}}},
\end{equation}
where $\overline{C^{2}} = \frac{\sum_{\ell=1}^{L}C_{\ell}^{2}}{L}$ is average squared \ac{SPPE} across districts, $\bar{k} = \frac{\sum_{i=1}^{N}k_{i}}{N}$ represents the average disutility from privacy loss across students, and $N$ is the number of students. The parameter $\eta = \phi/(1-\phi)$ measures the relative weight on privacy in the social welfare function.


\subsection{Solution and Calibration}
To establish a benchmark measure of WTA, we draw on the \ac{CCD} 2014---2015 \citep{CCD:2014}.
There are around $L=13,000$ public school districts and $N=46$ million school-age children.
Following \citet{Sonnenberg:TitleI:NCES:2016}, we calculate the adjusted \ac{SPPE} for each district. The average squared \ac{SPPE}, $\overline{C^{2}}$, is approximately 20 million.\footnote{Using the CCD, we deduct federal outlays from total state education expenditures and divide by average fall daily attendance to get unadjusted \ac{SPPE}. These are scaled down and truncated to get the adjusted \ac{SPPE} according to the process described by \citet{Sonnenberg:TitleI:NCES:2016}.
We then match each district to the adjusted \ac{SPPE} of its home state.
}

Completing the calibration requires information on preferences for privacy. In the \citet{Ghosh:Auction:GEB:2015} model, the privacy preference $k_{i}$ is a measure of the loss associated with states of the world that become more likely when data are published. We posit the cost of identity theft as a reasonable reference point. We therefore set $\bar{k} =\$1,400$ based on a Department of Justice estimate of the direct and indirect losses for victims of identity theft who lost at least one dollar \citep{Harrell:IdentityTheft:DOJ:2015}.\footnote{Technically, the estimated cost of identity theft reported by \citet{Harrell:IdentityTheft:DOJ:2015} is \$1,343. In keeping with our goal of offering a simple, stylized analysis, we round up to the nearest hundred dollars.} Hence, $WTA = \eta\times 1400\times 2.3 = \eta\times 3220$.

Setting $WTA = MRT$ and making all relevant substitutions,  $\varepsilon = 2.52\times \eta^{-\frac{1}{3}}$.
We report the cost of misallocation as the \ac{RMSE} in expected allocation across districts.\footnote{%
$RMSE = \sqrt{\mathbb{E}\left[L^{-1}\sum_{\ell=1}^{L}C^{2}_{\ell}\left(\widehat{E}_{\ell}-E_{\ell} \right)^{2}\right]}=\sqrt{-L^{-1}\overline{C^{2}}I}$.}
The \ac{RMSE} is measured in dollars per school district, so is comparable to other district-level expenditures.

We compare some benchmark models that place different relative weight on allocative efficiency. If privacy and accuracy are valued symmetrically, so $\eta=1$, the optimal level of privacy loss is $\varepsilon^{\ast}=2.52$, and the \ac{RMSE} in allocations across districts is approximately \$2,509. This reflects a misallocation cost of about 70 cents per student. One might favor greater allocative efficiency, since the cost of misallocation affects everyone in the population.
We might set $\eta = \frac{N}{POP-N}\approx 0.15$, where $POP$ is the total U.S.\ population. Then $\varepsilon^{\ast\ast} = 4.74$ and allocative inefficiency is just \$1,334, or approximately 38 cents per student.

Privacy advocates typically recommend values for $\varepsilon$ that are less than 1 and much closer to zero \citep{Dwork:Algorithmic:Book:2014}.
If we target $\varepsilon^{0}=0.1$, the \ac{RMSE} in allocations across districts is approximately \$63,000. At roughly 18 dollars per student, the same amount would cover the cost of lunch for seven days.\footnote{Based on a \$2.28 average cost per school lunch \citep{USDA:SchoolLunch:2008}.} The implied $\eta$ is around 12,000.

	\section{Application to Current Statistical Programs}
	\label{sec:applications}
%
%

Our analysis focuses on how statistical agencies can optimally choose publication strategies that balance privacy and accuracy in a disciplined manner. However, most statistical agencies are not yet using formal privacy protection systems. In this section, we describe how the tools and concepts developed in this paper may be brought to bear on several real-world problems confronting the U.S.\ statistical system.


\subsection{Legislative Redistricting}
\label{sec:redistricting_example}
The Census Bureau is required to provide geographically detailed counts of the population
to assist states in drawing legislative districts in compliance with the Equal Protection Clause of the $14^{th}$ Amendment of the U.S.\ Constitution and provisions of the 1965 Voting Rights Act.
These PL94-171 redistricting statistics
provide block-level population counts including data on race and ethnicity as mandated by the \citet{OMB:1997}.
The Census Bureau applies \ac{SDL} to these tabulations; however, the procedures used the 2000 and 2010 Census, and those proposed for the 2018 End-to-End Census Test, do not protect the counts of total or voting-age population at any level of geography, including the block.\footnote{See Appendix \ref{app:reconstruction_details} for the legislative and statutory background.}
In implementing the amendments to the Census Act, the Census Bureau has acted as if the social welfare function put all the weight on accuracy when drawing fresh legislative districts. However, it has given weight to privacy protection in publishing the statistics used to enforce the Voting Rights Act.

An attacker with the information set posited in equation (\ref{eq:semanticbound}) can always correctly determine the census block of the person missing from that information set.
Secret pairs like ``Bob lives on block 1'' versus ``Bob is not in the database'' cannot be protected whereas secret pairs like ``Bob is white (in the data)'' versus ``Bob left the race questions blank (in the data)'' can.
The absence of any provable privacy guarantee on the block-level geocode
means it is impossible to provide any meaningful privacy protection on the geolocation in any public-use tables or micro-data published from the decennial census. The block geocodes are ``blatantly non-private'' \cite[p. 204]{Dinur2003}.
Hence, releasing block-level population counts without privacy protection still affects the Census Bureau's ability to protect other variables.
The definition of neighboring databases must be changed. Only those matching the pre\-vi\-ous\-ly-published outputs are feasible.\footnote{The applicable variant is \emph{bounded} differential privacy \citep{Kifer:2011:NFL:1989323.1989345}.}
By publishing the PL94-171 tables with exact population counts in all geographies, the Census Bureau has constrained its ability to protect confidentiality---it always exposes the block-level geocode from the confidential data.

\subsection{Economic Censuses and National Accounts}
\label{sec:econ_census}
We have assumed the set of statistics to publish (the query workload) is pre-specified.
Producing tables with greater granularity reduces the accuracy of any specific table, holding privacy loss constant.
How to choose the query workload is another open question.

Every five years the Census Bureau conducts an economic census of business establishments.
A primary purpose is to provide statistical summaries against which the \ac{BEA} benchmarks the national accounts.
The \ac{BEA} benchmarks take as inputs the same detailed geographic, industrial, and product summaries that the Census Bureau releases to the general public. These publications are prepared using traditional \ac{SDL} methods. Regardless of the method, however, considerably more privacy protection is applied to these publications than would be required if the \ac{BEA} used the Census Bureau's confidential inputs, and then applied the privacy protections directly to the national account summaries before publication. Such a publication technology would have greater accuracy and better privacy protection---it would Pareto dominate the status quo. Although the Census Bureau and the \ac{BEA} are distinct agencies, \ac{CIPSEA} permits the exchange of confidential business data between these agencies. The U.S.\ could achieve at least some of the efficiency gains that countries with consolidated statistical agencies achieve by relying more on the authorized data sharing provisions of that law.

\subsection{Tax Data and Tax Simulations}
\label{sec:taxdata}
Formal privacy systems allow for other innovative approaches to data dissemination.
Consider the publication of data on individual, audited tax returns that are essential inputs for simulating the effects of tax policies \citep{feenberg:coutts:1993}.\footnote{The Congressional Budget Office, the Joint Committee on Taxation, the Office of Tax Analysis (Treasury), the Federal Reserve System, many research organizations, and statistical agencies run tax simulations.}
The validity of tax simulations depends on each input record reflecting the exact relationship between income and tax liability. The \ac{SOI} Division of the \ac{IRS} has regularly published a public-use micro-data file of tax returns protected by micro-aggregation in which the data from small sets of input records, usually three, were pooled together \citep[p. 49]{spwp22}.\footnote{The Statistics of Income Division currently sells these micro-data files, through 2012, for \$4,000 \citep{IRS:SOIProducts}. For documentation see \citet{NBER:GDB}.}
The micro-aggregation breaks some features of the audit consistency of the records and smooths over important breakpoints in the tax schedule, making them very hard to use in tax simulations.
To release data suitable for tax simulations using differentially private mechanisms would require very large $\varepsilon$, even though tax returns are highly sensitive.

A different type of formal privacy protection could address both the technology and social choice for the tax simulation problem. \ac{SOI} could run tax simulations behind its own firewall, using confidential tax returns as the inputs. The outputs of the tax simulation could be released using a differentially private publication system. Each user of the tax simulation system would have its own privacy-loss parameter, which \ac{SOI} could control for global accounting of the privacy loss associated with simulations. Since there is no requirement to hide the parameters of the differentially private publication system, the expected accuracy of the simulation outputs would be publicly known. These accuracy measures could then be used to make valid inferences about the effects of changing the inputs of the simulation. Errors induced by privacy protection could be considered in the same framework as other errors in the tax simulation. This solution to the social choice problem replaces an inefficient technology, traditional \ac{SDL}, with an efficient one, formal privacy. It permits \ac{SOI} to do the global accounting necessary to verify that its chosen weight on privacy protection relative to accuracy is properly implemented.\footnote{The system proposed here relies on a differentially private interactive query system, such as those described by \citet{Gupta2012}. 
}

\subsection{General Purpose Public-use Micro-data}
\label{sec:microdata_protection}
The Census Bureau published the first large-scale machine-readable \ac{PUMS} from the 1960 Census, selecting a one-percent sample of the records on the long form \citep{Ruggles:Frozen:HistMeth:2011}.
The goal of the \ac{PUMS} is to permit statistical analyses not supported by the published tabular statistics.
Generating formally private microdata is a daunting challenge. A better strategy may be to develop new privacy-preserving approaches to problems that have historically been solved by PUMS.

One approach is an online system that interactively answers requests for tabular and statistical models. All such requests can be thought of as queries. When an agency can specify the set of allowable queries in advance, it is possible to design a formally private publication mechanism that operates on the confidential micro-data and returns answers with known accuracy.
A formally private \ac{PUMS} would be dynamic, like the differentially private query systems embedded in Google's Chrome Browser, Apple's iOS 11, and Microsoft's Windows 10.\footnote{For the Chrome browser see \citet{DBLP:journals/corr/FantiPE15}. For iOS 11 see \citet{Apple:Learning:2017}. For Windows 10 see \citet{Ding:Telemetry:NIPS:2017}.}

An interactive online system works for models whose structure the agency can anticipate in advance (for example, the class of all linear regression models). More complicated analyses can be conducted in restricted-access environments. The Census Bureau has even acknowledged this publicly \citep{Census:Restricted:site}.
Restricted-access environments don't automatically protect privacy. The data custodian still needs to decide how much of the privacy-loss budget to reserve for such unstructured studies, and the engineers still need to build formally private data analysis systems.\footnote{Several such systems have been developed: see \citet{McSherry:PINQ:SIGMOD:2009}, \citet{chen:machanavajjhala:reiter:barrientos:IEEE:ICDM:2016}, and \citet{Harvard:DataPrivacyLab}.}


	\section{Directions for Research}
	\label{sec:conclusion}

This paper has developed a coherent and rigorous framework for guiding decisions about how to effectively publish population statistics while preserving privacy. To make our framework practical will require more sophisticated models of production possibilities as well as better models, and measures, of the demand for privacy and accuracy. We briefly consider several promising extensions.

\subsection{Extensions of the Production Model}
We have focused on publication of contingency tables and other population aggregates.
Our model of accuracy assumes that we only care about learning the finite-population statistic. This is not the same as learning about super-population parameters of the process that generated the data, as noted in Section \ref{sec:preferences}.
Our approach can be extended to statistical learning \citep{Wasserman:Statistical:JASA:2010,Duchi:Minimax:IEEE:2013,DBLP:journals/corr/DworkR16}.
However, a robust approach must also allow for ambiguity regarding the true data generating process.

Our analysis of the economic census in Section \ref{sec:econ_census} suggests that the set of statistics to publish should be endogenous.
Doing so requires a production technology that allows for different analyses to have different accuracy guarantees under the same publication system.
For example, one could endow different users with a fixed privacy-loss budget and let them submit queries. The answers would be accurate in proportion to the ratio of the query sensitivity to the privacy-loss endowment. More broadly, decisions about data collection, the set of tables to publish, and the privacy-loss budget, should be determined simultaneously.
Our paper is a necessary step on the way toward a comprehensive analysis of decisions about collection, processing, and dissemination of information by statistical agencies.

We assume the statistical agency can explicitly enumerate its feasible combinations of privacy loss and statistical accuracy. This works when the statistical agency uses differentially private mechanisms with known accuracy and privacy guarantees. In more realistic settings where data are also subject to complex editing constraints, determining the production function is challenging. This is an active area of research.

\subsection{Extensions of the Model of Preferences}
The idea that statistical accuracy and privacy protection are public goods is not controversial, but does not often appear in models of data provision. We need models of market provision of statistics when those summaries are public goods. Such a model might start by extending the framework posed by \citet{Ghosh:Auction:GEB:2015} along the lines of \citet{Spence:Monopoly:Bell:1975}, noting that those who sell their data value privacy less than the marginal consumer whose preferences a monopolistic provider will internalize.\footnote{We are grateful to an anonymous referee for pointing out this extension.}

As \citet{Nissim:2012:PMD:2229012.2229073} point out, differential privacy only bounds the worst-case harm from participation. As one approach toward developing a better model, \citet{Kifer2012} suggest building data security algorithms based directly on privacy semantics. Turning to preferences for accuracy, we have assumed a reduced-form relationship between public statistics and wealth. A deeper analysis would directly model the role of public statistics in improving decision-making, as in \citet{Spencer:Optimal:JASA:1985}. Alternatively, public statistics could enter production as a form of public capital, with accuracy affecting consumption through the demand for labor or reduced goods prices.

While differential privacy guarantees do not change over time, our static model abstracts from the possibility that the costs of privacy loss and the benefits of accurate statistics might time-vary.
The Census Bureau's practice of making the complete responses from each decennial census public after 72 years is an implicit acknowledgment that privacy preferences (or, equivalently, the costs of privacy loss) change over time, at least relative to the social benefit of access to the full data. Further study is needed to determine how these dynamic considerations should factor into our basic social choice framework.

Finally, our models of data collection implicitly assume truthful reporting. With declining rates of survey participation, understanding the connection between data publication and data collection is also important. New thinking about who creates and who buys data, proposed in \citet{arrieta:dataAsLabor:2017}, can also inform models of data acquisition by statistical agencies. \citet{Nissim:2012:PMD:2229012.2229073}, \citet{Xiao:Privacy:ITCS:2013}, and \citet{Chen:2016:Truthful:2016} also study the problem of eliciting truthful responses in the presence of privacy concerns.

\subsection{The Need for Better Measurement}
We need to learn more about preferences for privacy and accuracy.
There is a growing body of evidence from public opinion surveys on attitudes toward privacy \citep{Childs:Development:JSM:2012, childs:trust:2014, Childs:Confidence:SP:2015}.
While these are instructive, it is far from clear how reported attitudes correspond to behavioral responses associated with changes in the risk of privacy loss.
Some experiments have informed the value people attach to providing private data for commercial use \citep{Acquisti2013}.
More information is needed on the price people attach to privacy loss, particularly as regards the inferential disclosures considered in this paper.
With few exceptions \citep{Spencer:Optimal:JASA:1985,Mulry:Accuracy:JASA:1990,Seeskin:Spencer:Effects:2015}, there is virtually no evidence on the social value of public statistics, let alone the value of improving their accuracy.

\subsection{Conclusion}
Formal privacy models facilitate an interpretation of privacy protection as a commodity over which individuals have preferences. When statistical accuracy and privacy are public goods, as is the case for the publication of official statistics, their optimal levels are a social choice.
This social choice can, and should, be guided by the principle of equating marginal social costs with marginal social benefits.
In developing these ideas, we made many simplifying assumptions. We hope these can be excused as a feature of combining insights from three different disciplines to bear on a question of substantial scientific and public interest. We also hope our paper motivates researchers from economics, demography, computer science, statistics, and related disciplines to take up the Information Age challenge of designing publication systems that support accurate science but do not require infinite privacy loss.





  \begin{center}
{\Large\bf \mytitle \ \\ \ \\ John M.\ Abowd and Ian M.\ Schmutte} \ \\ \ \\ {\bf Online Appendix   \ \\ \ \\ \today}
\end{center}

	\pagenumbering{arabic}\renewcommand{\thepage}{App.~\arabic{page}}
	\setcounter{section}{0} \renewcommand{\thesection}{\Alph{section}}
	\renewcommand{\thesubsection}{\Alph{section}\arabic{subsection}}
	\setcounter{equation}{0} \renewcommand{\theequation}{\Alph{section}-\arabic{equation}}
	\label{sec:appendix}
%
%

\section{Potential Database Reconstruction Attack against the 2010 Decennial Census}
\label{app:reconstruction_details}
In Section \ref{sec:reconstruction} we discuss the potential for a database reconstruction attack against the decennial census based on the large number of summary tables published from the confidential micro-data. Using the schema in the public documentation for PL94-171, Summary File 1, Summary File 2, and the Public-use Micro-data Sample, and summarizing from the published tables, there were at least 2.8 billion \emph{linearly independent} statistics in PL94-171, 2.8 billion in the balance of SF1, 2.1 billion in SF2, and 31 million in the PUMS \url{https://www.census.gov/prod/www/decennial.html} (cited on March 17, 2018).
For the 2010 Census, the national sample space at the person level has approximately 500,000 cells. The unrestricted sample space at the census block level has approximately 500,000 $\times$ $10^7$ cells. It might seem there are orders of magnitude more unknowns than equations in the system used for reconstruction. However, traditional \ac{SDL} does not protect sample zeros. Consequently, every zero in a block, tract, or county-level table rules out all record images in the sample space that could have populated that cell, dramatically reducing the number of unknowns in the relevant equation system.

The deliberate preservation of sample zeros can be inferred from the technical documentation: ``Data swapping is a method of disclosure avoidance designed to protect confidentiality in tables of frequency data (the number or percentage of the population with certain characteristics). Data swapping is done by editing the source data or exchanging records for a sample of cases. A sample of households is selected and matched on a set of selected key variables with households in neighboring geographic areas (geographic areas with a small population) that have similar characteristics (same number of adults, same number of children, etc.). Because the swap often occurs within a geographic area with a small population, there is no effect on the marginal totals for the geographic area with a small population or for totals that include data from multiple geographic areas with small populations. Because of data swapping, users should not assume that tables with cells having a value of one or two reveal information about specific individuals'' \citep[p. 7-6]{sf1:2010}.

\section{Randomized Response Details}
\label{app:randomized_response}
A custodian collects data from a population of individuals, $i\in\left\{1,\ldots,N\right\}$.
Each member of the population has a sensitive characteristic and an innocuous characteristic.
The sensitive characteristic is $x_{i} = Y_{i}(1)\in\left\{0,1 \right\}$, with population proportion $\PrP\left[Y_{i}(1)=1 \right] = \pi$.
This proportion, $\pi$, is the unknown population quantity of interest.
The non-sensitive characteristic is $z_{i} = Y_{i}(0)\in\left\{0,1 \right\}$ with known population proportion $\PrP\left[Y_{i}(0)=1 \right] = \mu$.
The custodian collects and publishes a mixture
\begin{equation}
	d_{i} = T_{i}Y_{i}(1) + (1-T_{i})Y_{i}(0),
	\label{eq:RR_var}
\end{equation}
where $T_{i}$ indicates whether the sensitive or the non-sensitive question was collected, with $\PrP\left[T_{i}=1 \right] = \rho$.
The responses are independent of which information is collected: $\left(Y_{i}(1),Y_{i}(0)\right)\Perp T_{i}$. We also require that the non-sensitive item be independent of the sensitive item. This is not restrictive, since the innocuous question can literally be ``flip a coin and report whether it came up heads,'' as in the original application.

The indicator $T_{i}$ is not observed. Any data analyst observes only the reported variable $d_{i}$.
However, as in a randomized controlled trial, the probability of $T_{i}$, $\rho$, is known with certainty. Furthermore, the analyst also knows the probability of the non-sensitive response, $\mu$.

Define $\widehat{\beta} = \frac{1}{N}\sum_{i}d_{i}$, the empirical mean proportion of responses of one.
Independence of $T_{i}$ implies $\E\left[\widehat{\beta}\right] = \pi\rho + \mu(1-\rho)$.
It follows that $\widehat{\pi} = \frac{\widehat{\beta}-\mu(1-\rho)}{\rho}$ is an unbiased estimator of $\pi$ with variance $\limfunc{Var}[{\widehat{\pi}}] = \limfunc{Var}[{\widehat{\beta}}]\rho^{-2}$.

\subsection{Privacy under Randomized Response}

For given $\varepsilon$, differential privacy requires both
$\PrP\left[d_{i}=1|Y_{i}(1)=1 \right]$ $\leq$ $e^{\varepsilon}\PrP\left[d_{i}=1|Y_{i}(1)=0 \right]$, and $\PrP\left[d_{i}=0|Y_{i}(1)=0 \right]$ $\leq$ $e^{\varepsilon}\PrP\left[d_{i}=0|Y_{i}(1)=1 \right]$. Together, these expressions bound the Bayes factor, which limits how much can learned about the sensitive characteristic upon observation of the collected response.

Making substitutions based on the data-generating model,
\begin{equation}
	1 + \frac{\rho}{(1-\rho)\mu}\leq e^{\varepsilon}
	\label{eq:dp_bayesfactor_1}
\end{equation}
and
\begin{equation}
	1 + \frac{\rho}{(1-\rho)(1-\mu)}\leq e^{\varepsilon}.
	\label{eq:dp_bayesfactor_2}
\end{equation}
For a given choice of $\mu$, the differential privacy guaranteed by randomized response is the maximum of the values of the left-hand sides of equations \eqref{eq:dp_bayesfactor_1} and \eqref{eq:dp_bayesfactor_2}. Hence, privacy loss is minimized when $\mu=\frac{1}{2}$. This is the case we will consider throughout the remaining discussion. We note doing so assumes that inferences about affirmative and negative responses are equally sensitive, which may not always be the case. The results of our analysis do not depend on this assumption. \footnote{These observations highlight another allocation problem: how to trade off protection of affirmative responses for the sensitive item $Y_{i}(1)=1$ against protection of negative responses $Y_{i}(1)=0$.
What do we mean? If $\rho$ is fixed, then increasing $\mu$ reduces the Bayes factor in \eqref{eq:dp_bayesfactor_1} (increasing privacy) and increases the Bayes factor in \eqref{eq:dp_bayesfactor_2} (decreasing privacy). The underlying intuition is fairly simple. Suppose the sensitive question is ``did you lie on your taxes last year?'' Most tax evaders would prefer that their answer not be made public, but non-evaders are probably happy to let the world know they did not cheat on their taxes. In such a setting, with $\rho$ fixed, we can maximize privacy for the tax evader by setting $\mu$ to 1. Recall $\mu$ is the probability of a positive response on the non-sensitive item ($Y_{i}(0)=1$). If $\mu=1$, then when the data report $d_{i}=0, 1$ we know with certainty that $Y_{i}(1)=0$ (i.e., $i$ did not cheat on her taxes). In this special case, the mechanism provides no privacy against inference regarding non-evasion, but maximum attainable privacy (given the mechanism) against inference regarding evasion. This is the role the Bloom filter plays in the full RAPPOR implementation of randomized response \citep{Erlingsson2014}. More generally, the choice of $\mu$ can be tuned to provide relatively more or less privacy against one inference or the other.}

For randomized response, the differential privacy guarantee as a function of $\rho$ is:
\begin{equation}
 	\varepsilon(\rho) = \log\left(1 + \frac{2\rho}{(1-\rho)}\right)=\log\left(\frac{1+\rho}{1-\rho} \right),
 	\label{eq:rand_response_epsilon_func_app}
 \end{equation}
 which follows from setting $\mu=\frac{1}{2}$ in equations \eqref{eq:dp_bayesfactor_1} and \eqref{eq:dp_bayesfactor_2}.

\subsection{Statistical Accuracy under Randomized Response}
Expressed as a function of $\rho$, we denote the estimated share of the population with the sensitive characteristic
\begin{equation}
	\widehat{\pi}\left( \rho\right) = \frac{\widehat{\beta}(\rho)-\mu(1-\rho)}{\rho}
\end{equation}
where $\widehat{\beta}(\rho)$ is the population average response when the sensitive question is asked with probability $\rho$. Clearly,
\begin{equation}
	\E[\widehat{\beta}(\rho)] = [\rho\pi + (1-\rho)\mu]
\end{equation}
and
\begin{equation}
	\limfunc{Var}[\widehat{\beta}(\rho)] = \frac{[\rho(\pi-\mu) + \mu](1-\rho(\pi-\mu)-\mu)}{N}.
\end{equation}
It follows that
\begin{equation}
	\limfunc{Var}[\widehat{\pi}(\rho)] = \frac{\limfunc{Var}[\widehat{\beta}(\rho)]}{\rho^{2}} = \frac{[\rho(\pi-\mu) + \mu](1-\rho(\pi-\mu)-\mu)}{\rho^{2}N}.
\end{equation}


We can define data quality as:
\begin{equation}
 	I(\rho) = \limfunc{Var}[\widehat{\pi}\left( 1\right)] - \limfunc{Var}[\widehat{\pi}\left( \rho\right)] .
 	\label{eq:rand_response_dataquality_func}
\end{equation}
This measures the deviation in the sampling variance for the predicted population parameter, $\pi$, relative to the case where there is no privacy protection ($\rho=1$).

\subsection{The Accuracy Cost of Enhanced Privacy under Randomized Response}

Equations \eqref{eq:rand_response_epsilon_func_app} and \eqref{eq:rand_response_dataquality_func} implicitly define a functional relationship between data privacy, parameterized by $\varepsilon$, and accuracy, parameterized as $I$. This function tells us the marginal cost borne by individuals in the database necessary to achieve an increase in accuracy of the published statistics.
We can characterize the relationship between accuracy, $I$, and privacy loss, $\varepsilon$, analytically. First, we invert equation \eqref{eq:rand_response_epsilon_func_app} to get $\rho$ as a function of $\varepsilon$:
\begin{equation}
 	\rho(\varepsilon) = \frac{e^{\varepsilon}-1}{1+e^{\varepsilon}}.
\end{equation}
Next, we differentiate $I$ with respect to $\varepsilon$ via the chain rule: $\frac{dI}{d\varepsilon} = I^{\prime}(\rho(\varepsilon))\rho^{\prime}(\varepsilon)$:
\begin{equation}
 	I^{\prime}(\rho) = \frac{2\limfunc{Var}[\widehat{\beta}(\rho)]}{\rho} - \frac{(\pi-\frac{1}{2})(1-2\pi)}{N^{2}\rho}.
\end{equation}
and
\begin{equation}
 	\rho^{\prime}(\varepsilon) = \frac{2e^{\varepsilon}}{(1+e^{\varepsilon})^{2}}=\frac{1}{1+cosh(\varepsilon)}.
\end{equation}
Both derivatives are positive, so it follows that $\frac{dI}{d\varepsilon}>0$. A similar derivation shows that $\frac{d^{2}I}{d\varepsilon^{2}}<0$. Increasing published accuracy requires an increase in privacy loss at a rate given by $\frac{dI}{d\varepsilon}>0$. Furthermore, achieving a given increment in accuracy requires increasingly large privacy losses. 

\section{Details of the Matrix Mechanism}
\label{app:matrix_mechanism}
 For a single query, we defined the $\ell_{1}$ sensitivity in Definition \ref{def:query_sensitivity}. The results in Theorems \ref{prop:laplace_mechanism} and \ref{thm:ppf} are defined in terms of the sensitivity of a workload of linear queries, which we denote $\Delta Q$. Following \citet{li:matrix:VLDB:2015},
\begin{theorem} [$\ell_{1}$ Query Matrix Sensitivity]
\label{prop:query_workload_sensitivity}
Define the $\ell_{1}$ sensitivity of $Q$ by
$$\Delta Q = \max_{x,y\in \mathbb{Z}^{\ast |\chi |}, ||x-y||_{1} \le 1}\left\Vert Qx - Qy\right\Vert_{1}.$$
This is equivalent to
\begin{equation*}
	 \Delta Q = \limfunc{max}_{k}\left\Vert q_{k}\right\Vert_{1},
\end{equation*}
where $q_k$ are the columns of $Q$.
\end{theorem}
For the proof, see \citet[prop. 4]{li:matrix:VLDB:2015}.

\section{Details of Privacy Semantics}
\label{app:dp_semantics_details}
We provide technical definitions associated with the derivations in \citet{Kifer2012} described in Section \ref{sec:dp_semantics}.

Assume a latent population of individuals $h_{i}\in\mathcal{H}$ of size $N^*$.
The confidential database, $D$, is a random selection of $N < N^*$ individuals, drawn independently from $\mathcal{H}$.
In this context $N$ is a random variable, too.
The database also records characteristics of each individual, which are drawn from the data domain $\chi$.
Denote the record of individual $i$ as $r_i$.
The event ``the record $r_i$ is included in database $D$'' has probability $\pi_i$.
Denote the conditional probability of the event ``the record $r_i=\chi_{a} \in \chi$'' given that $r_i$ is in $D$ as $f_{i}(r_{i})$.
Then, the data generating process is parameterized by $\theta = \{\pi_1, ..., \pi_N,f_1, ..., f_N\}$.
The probability of database $D$, given $\theta$, is
\begin{equation}
  Pr\left[D\ |\ \theta\right] = \prod_{h_i \in D}f_{i}(r_{i})\pi_i \prod_{h_j \notin D}(1-\pi_j).
  \label{eq:dgp}
\end{equation}

The complete set of paired hypotheses that differential privacy protects is
\begin{equation}
  \mathcal{S}_{pairs} = \{(s_{i}, s_{i}^{\prime}) : h_i \in \mathcal{H}, \chi_{a} \in \chi\},
  \label{eq:secretpair}
\end{equation}
where $s$ and $s^{\prime}$ are defined in Section \ref{sec:dp_semantics}. By construction $\mathcal{S}_{pairs}$ contains every pair of hypotheses that constitute a potential disclosure; that is, whether any individual $h_{i}$ from the latent population is in or out of the database $D$ and, if in $D$, has record $r_i$.

\section{Derivation of the Data Utility Model}
\label{app:data_utility_details}
Recall that the matrix mechanism publishes a vector of answers, $M(x,Q)$ to the known set of queries, $Q$ given an underlying data histogram $x$. The matrix mechanism is implemented by using a data independent mechanism to answer a set of queries represented by the query strategy matrix, $A$ with sensitivity $\Delta A$ and pseudo-inverse $A^{+}$. Following Theorem \ref{thm:ppf}, $M(x,Q) = Qx + Q(\Delta A)A^{+}e$ where $e$ is a vector of $iid$  random variables with $\mathbb{E}\left[e\right]=0$ and whose distribution is independent of $x$, $Q$, and $A$. In what follows, we use the notation $\sigma^{2}_{e}$ to denote the common (scalar) variance of the elements of $e$. For example, when $e$ is a vector of Laplace random variables with scale $\varepsilon^{-1}$, we know that $\sigma^{2}_{e} = 2\varepsilon^{-2}$. Note that the variance of the vector $e$ is $\mathbb{E}\left[ee^{T}\right] = \sigma^{2}_{e} \mathbb{I}$ where $\mathbb{I}$ is the identity matrix conformable with $e$.

Let $W_{i} = \Pi_{i}^{T} M(x,Q)$ be a person-specific linear function by which published statistics are transformed into wealth (or consumption). Individuals have utility of wealth given by a twice-differentiable and strictly concave function, $U_{i}(W_{i})$.
The total realized ex post wealth for $i$ is
$
	W_{i} = \Pi_{i}^{T}Qx + \Pi_{i}^{T}QA^{+}(\Delta A)e.
$
We assume $i$ knows $Q$ and the details of the mechanism $M$. Uncertainty is over $x$ and $e$. 

For notational convenience, we define a function $w_{i}(e;x) = \Pi_{i}^{T}Qx + \Pi_{i}^{T}QA^{+}(\Delta A)e $.
Conditional on $x$, the expected utility of $i$ from receiving the mechanism output is 
$
	\mathbb{E}_{e|x}\left[U_{i}\left(w_{i}(e;x)\right)\vert x\right]
$.
We approximate this by taking expectations of a second-order Taylor Series expansion of $h_{i}(e;x) = U_{i}\left(w_{i}(e;x)\right)$ with respect to $e$ evaluated at $e_{0}=0$.

Let $\nabla h_{i}(e_{0};x)$ denote the gradient of $h$ with respect to $e$ and let $H_{i}(e_{0};x)$ denote the Hessian. The second-order Taylor series expansion of $h_{i}(e;x)$ evaluated at $e_{0}$ is
\begin{equation}
	h_{i}(e;x) \approx h_{i}(e_{0};x) + (e - e_{0})^{T}\nabla h_{i}(e_{0};x) + \frac{1}{2!} (e - e_{0})^{T} H_{i}(e_{0};x) (e - e_{0}).
	\label{eq:taylor_expansion_general}
\end{equation}
The gradient of $h$ is
\begin{equation}
	\nabla h_{i}(e_{0};x) = U_{i}^{\prime}(w_{i}(e_{0};x)) \Delta A \left(\Pi_{i}^{T}QA^{+}\right)^{T}.
\end{equation}
The Hessian is
\begin{equation}
	H_{i}(e_{0};x) = U_{i}^{\prime\prime}(w_{i}(e_{0};x)) (\Delta A)^{2}\left(\Pi_{i}^{T}QA^{+}\right)^{T}\left(\Pi_{i}^{T}QA^{+}\right).
\end{equation}
Note that we have used the chain rule in both derivations.
We now evaluate the right hand side of equation \eqref{eq:taylor_expansion_general} at $e_{0}=0$. Defining new notation, let $w_{i0}^{x} = w_{i}(0;x) = \Pi_{i}^{T}Qx $ and making substitutions for the gradient and Hessian, we have
\begin{align}
	h_{i}(e;x) \approx U_{i}\left(w_{i0}^{x}\right) + U_{i}^{\prime}(w_{i0}^{x}) \Delta A \left[e^{T}\left(\Pi_{i}^{T}QA^{+}\right)^{T}\right] + \frac{1}{2}U_{i}^{\prime\prime}(w_{i0}^{x}) \Delta A^{2} \left[e^{T}\left(\Pi_{i}^{T}QA^{+}\right)^{T}\left(\Pi_{i}^{T}QA^{+}\right)e\right].
\end{align}
Now, taking expections with respect to $e$, conditional on $x$
\begin{equation}
	\mathbb{E}_{e|x}\left[h(e;x)\vert x\right] \approx U_{i}\left(w_{i0}^{x}\right) + \frac{1}{2}U_{i}^{\prime\prime}(w_{i0}^{x}) \Delta A^{2} \cdot \mathbb{E}_{e|x}\left\{ \left[e^{T}\left(\Pi_{i}^{T}QA^{+}\right)^{T}\left(\Pi_{i}^{T}QA^{+}\right)e\right] \vert x\right\}.
\end{equation}
The first-order term drops out because $\mathbb{E}_{e|x}\left[e\vert x \right] = 0$ by assumption. Focusing on the quadratic form in the final summand, standard results imply
\begin{align}
	\mathbb{E}_{e|x}\left\{ \left[e^{T}\left(\Pi_{i}^{T}QA^{+}\right)^{T}\left(\Pi_{i}^{T}QA^{+}\right)e\right] \vert x\right\} & = tr\left[\mathbb{E}_{e|x}\left[ee^{T}\vert x\right] \left(\Pi_{i}^{T}QA^{+}\right)^{T}\left(\Pi_{i}^{T}QA^{+}\right) \right] \\
	& = tr\left[\sigma_{e}^{2} \mathbb{I} \left(\Pi_{i}^{T}QA^{+}\right)^{T}\left(\Pi_{i}^{T}QA^{+}\right) \right] \\
	& = \sigma_{e}^{2} tr\left[\left(\Pi_{i}^{T}QA^{+}\right)^{T}\left(\Pi_{i}^{T}QA^{+}\right) \right] \\
	& = \sigma_{e}^{2} \Vert \Pi_{i}^{T} QA \Vert ^{2}_{F} . 
\end{align}
The last expression is a basic property of the matrix Frobenius norm \citep{li:matrix:VLDB:2015}.

Putting it all together, we have the following approximation to the expected utility for person $i$:
\begin{align}
	\mathbb{E}[U_{i}(W_{i})] 
	& = \mathbb{E}_{x}\left[\mathbb{E}_{e|x}\left[ h(e;x) \vert x \right] \right] \\
    & \approx \mathbb{E}_{x}\left[U_{i}\left(w_{i0}^{x}\right) + \frac{1}{2}U_{i}^{\prime\prime}(w_{i0}^{x}) \Delta A^{2} \sigma_{e}^{2} \Vert \Pi_{i}^{T} QA \Vert ^{2}_{F}\right] \\
    & =  \mathbb{E}_{x}\left[U_{i}\left(w_{i0}^{x}\right)\right] + \frac{1}{2} \mathbb{E}_{x}\left[U_{i}^{\prime\prime}(w_{i0}^{x})\right]\Delta A^{2} \sigma_{e}^{2} \Vert \Pi_{i}^{T} QA \Vert ^{2}_{F}.
\end{align}
Note that we have used the fact that $A$, $Q$, and $\Pi_{i}^{T}$ are all independent of $x$.

From Theorem \ref{thm:ppf} the accuracy of the matrix mechanism is
\begin{equation}
	I = - \sigma^{2}_{e} \left(\Delta A\right)^{2}\left\Vert QA^{+}\right\Vert_{F}^{2}.
\end{equation}
We can therefore substitute accuracy, $I$, into the expression for expected utility
\begin{equation}
	\mathbb{E}[U_{i}(W_{i})] \approx  \mathbb{E}_{x}\left[U_{i}\left(w_{i0}^{x}\right)\right] -  \left\{\frac{1}{2}\mathbb{E}_{x}\left[U_{i}^{\prime\prime}(w_{i0}^{x})\right] \frac{\Vert \Pi_{i}^{T} QA \Vert ^{2}_{F}}{\Vert QA \Vert ^{2}_{F}}\right\}\times I.
\end{equation}
The expression above rationalizes a model for individual-specific data utility that is linear in accuracy, $I$: $v^{Data}_{i}(I) = a_{i} + b_{i} I$.

\section{Details of Legislative Redistricting Example}
This appendix describes the legal background for the legislative redistricting example in Section \ref{sec:redistricting_example}.
These properties of the \ac{SDL} applied in the 2010 PL94-171 can be deduced from \citet[p. 7-6]{sf1:2010}, as quoted in Appendix \ref{app:reconstruction_details}, and the details provided in \citet{Census:2002}, which also reveals that no privacy protection was given to the race and ethnicity tables in the 1990 redistricting data. The origin of the decision not to protect population and voting-age population counts is difficult to trace in the law. Public Law 105–119, title II, § 209, Nov. 26, 1997, 111 Stat. 2480, amended 13 U.S.C. Section 141 to provide that: ``(h) ... In both the 2000 decennial census, and any dress rehearsal or other simulation made in preparation for the 2000 decennial census, the number of persons enumerated without using statistical methods must be publicly available for all levels of census geography which are being released by the Bureau of the Census for: (1) all data releases before January 1, 2001; (2) the data contained in the 2000 decennial census Public Law 94–171 [amending this section] data file released for use in redistricting; (3) the Summary Tabulation File One (STF–1) for the 2000 decennial census; and (4) the official populations of the States transmitted from the Secretary of Commerce through the President to the Clerk of the House used to reapportion the districts of the House among the States as a result of the 2000 decennial census. ... . (k) This section shall apply in fiscal year 1998 and succeeding fiscal years.'' \url{http://www.law.cornell.edu/uscode/text/13} \cite{title13}. These amendments to Title 13 concerned the use of sampling to adjust the population counts within states, as is permitted even under current law. They gave standing to obtain a copy of population count data that were not adjusted by sampling, should the Census Bureau publish such data, which it did not do in 2000 nor 2010. Even so, only the reapportionment of the House of Representatives must be done without sampling adjustments \citep{Commerce:v:House:1999}. Sampling aside, other statistical methods, like edits and imputations, including whole-person substitutions, are routinely applied to the confidential enumeration data before any tabulations are made, including those used to reapportion the House of Representatives. These methods were upheld in Utah v. Evans \citep{Utah:v:Evans:2002}.





	
\end{document}